\documentclass[structabstract]{aa}  
\usepackage{graphicx}
\usepackage{natbib}
\usepackage{txfonts}
\usepackage{color}
\newcommand\Zmax{$Z_\mathrm{max}$}

\newcommand\R{$R$}

\newcommand\Rg{$R_\mathrm{g}$}
\newcommand\temp{$T_{\fontsize{6}{6}\selectfont \mbox{eff}}$}\normalfont
\newcommand\logg{$\log g$}
\newcommand\met{[M/H]}

\begin{document}
   \title{Chemical gradients in the Milky Way from the RAVE data. I. Dwarf
 stars.}

   \author{
	C. Boeche\inst{1,2},  A. Siebert\inst{3}, 
	T. Piffl\inst{2}, A. Just\inst{1}, M. Steinmetz\inst{2}, 
	S. Sharma\inst{4}, G. Kordopatis\inst{5}, 
	G. Gilmore\inst{5}, C. Chiappini\inst{2}, 
	M. Williams\inst{2}, E. K. Grebel\inst{1}, 
	J. Bland-Hawthorn\inst{4}, B. K. Gibson\inst{6,7},
	U. Munari\inst{8}, 
	A. Siviero\inst{9,2}, O. Bienaym\'e\inst{3},
	J.F. Navarro\inst{10}, 
	Q. A. Parker\inst{11,12,13}, W. Reid\inst{12},
	G. M. Seabroke\inst{14},
	F. G. Watson\inst{13}, 
	R. F. G. Wyse\inst{15},
	 T. Zwitter\inst{16,17}
	}          

   \offprints{corrado@ari.uni-heidelberg.de}

   \institute{
Astronomisches Rechen-Institut, Zentrum f\"ur Astronomie der 
Universit\"at Heidelberg, M\"onchhofstr. 12-14, D-69120 Heidelberg
         \and
Leibniz Institut f\"ur Astrophysik Potsdam (AIP), An der Sternwarte 16, 
D-14482 Potsdam, Germany
         \and
Observatoire de Strasbourg, Universit\'e de Strasbourg, CNRS
11 rue de l'universit\'e, F-67000 Strasbourg, France
	\and
Sydney Institute for Astronomy, School of Physics A28, University of Sydney, NSW 2006, Australia
	\and
Institute of Astronomy, University of Cambridge, Madingley Road, Cambridge CB3 0HA, UK
        \and
Jeremiah Horrocks Institute, University of Central Lancashire, Preston,
 PR1~2HE, UK
        \and
Monash Centre for Astrophysics, School of Mathematical Sciences, Monash
 University, Clayton, VIC, 3800, Australia
	\and
INAF Osservatorio Astronomico di Padova, Via dell'Osservatorio 8, Asiago I-36012, Italy
        \and
Department of Physics and Astronomy, Padova University, Vicolo
dell'Osservatorio 2, I-35122 Padova, Italy
	\and
University of Victoria, P.O. Box 3055, Station CSC, Victoria, BC V8W 3P6, Canada
	\and
Department of Physics \& Astronomy, Macquarie University, Sydney, NSW 2109
 Australia 
	\and
Research Centre for Astronomy, Astrophysics and
Astrophotonics, Macquarie University, Sydney, NSW 2109 Australia 
	\and
Australian Astronomical Observatory, PO Box 915, North Ryde, NSW 1670, Australia
        \and
Mullard Space Science Laboratory, University College London, Holmbury St
Mary, Dorking, RH5 6NT, UK
        \and
Department of Physics and Astronomy, Johns Hopkins University, 3400 North
Charles Street, Baltimore, MD 21218, USA
        \and
Faculty of Mathematics and Physics, University of Ljubljana, Jadranska 19, SI-1000 Ljubljana, Slovenia
        \and
Center of Excellence SPACE-SI, Askerceva cesta 12, SI-1000 Ljubljana, Slovenia
}


 
  \abstract
  {}
  {We aim at measuring the chemical gradients of the elements Mg, Al, Si, 
  and Fe along the Galactic radius to provide new constraints on the chemical
  evolution models of the Galaxy and Galaxy models such as the Besan\c con
  model.  Thanks to the large number of stars of our RAVE sample we can study
  how the gradients vary as function of the distance from the Galactic plane.}
  {We analysed three different samples selected from three independent
  datasets: a sample of 19\,962 dwarf stars selected from the RAVE database, a
  sample of 10\,616 dwarf stars selected from the Geneva-Copenhagen Survey
  (GCS) dataset, and a mock sample (equivalent to the RAVE sample)
  created by using the GALAXIA code, which is based on the Besan\c con model.  
We integrated the Galactic orbits and obtained the
  guiding radii (\Rg) and the maximum distances from the Galactic plane reached
  by the stars along their orbits (\Zmax). We measured the chemical
  gradients as functions of \Rg\ at different \Zmax.}
  {The chemical gradients of the RAVE and GCS samples are
  negative and show consistent trends, although they are not equal: at
  \Zmax$<$0.4~kpc and 4.5$<$\Rg(kpc)$<$9.5, the iron gradient 
  for the RAVE sample is $d[Fe/H]/dR_\mathrm{g}=-0.065$~dex~kpc$^{-1}$, 
  whereas for the GCS sample it is 
  $d[Fe/H]/dR_\mathrm{g}=-0.043$~dex~kpc$^{-1}$ with internal errors
  of $\pm0.002$ and $\pm0.004$~dex~kpc$^{-1}$, respectively.
The gradients of the RAVE and GCS samples become flatter at larger \Zmax. 
Conversely, the mock sample has a positive iron gradient of
$d[Fe/H]/dR_\mathrm{g}=+0.053\pm0.003$~dex~kpc$^{-1}$ at \Zmax$<$0.4~kpc and
remains positive at any \Zmax.  These positive and unrealistic values originate
from the lack of correlation between metallicity and tangential velocity in the
Besan\c con model.
The discrepancies between the observational samples and the
mock sample can be reduced by i) decreasing the density, ii) decreasing the
vertical velocity, and iii) increasing the metallicity of the thick disc in
the Besan\c con model.}
   {}


   \titlerunning{RAVE chemical gradients in the MW}
    \authorrunning{Boeche et al.}

   \maketitle

%

\section{Introduction}\label{intro_grad}
The spatial distribution of the chemical abundances in the Milky Way
provides important constraints for our understanding of galaxy formation.
The distribution of the chemical species is a tracer of the way the galaxies
formed because the abundances in the stellar atmospheres of a stellar 
population do not change during its lifetime\footnote{An exception to this
occurs in the last stages of the stellar evolution, when the chemical
abundances of the stellar atmosphere can be changed by dredge-ups, which bring
the products of the nucleosynthesis to the surface.},
in contrast to its spatial distribution and kinematics.
Stars inherit the metallicity of the interstellar medium from which they
form. Its metallicity  
depends in a complex way on the infall and outflow of gas and the
enrichment by older stellar populations. Additionally, the element abundance
ratios depend on the speed of enrichment. Fast enrichment results in an
$\alpha$-enhancement because $\alpha$-elements are mainly produced by
supernovae (SNe) type II on a short timescale, whereas most of the iron is produced
by SN type Ia with a significant time delay. Therefore the abundances
of stellar populations are tightly connected to the gas infall and star
formation history. As a consequence, two locations that experienced
different star formation histories exhibit a difference (gradient) 
in chemical abundances.\\

Today the existence of a negative chemical gradient along the Galactic radius
is well established thanks to several dedicated observational works (see
Maciel \& Costa \citealp{maciel2010} and references therein).
Chemical gradients were measured by means of several tracers (such as
Cepheids, planetary nebulae, open clusters) that cover a large
range of Galactocentric distances (\R).
The chemical gradients found by previous investigations can span quite a
wide range (from $-0.029$~dex~kpc$^{-1}$ by Andrievsky et al., 
\citealp{andrievsky} using Cepheids, to $-0.17$~dex~kpc$^{-1}$ by Sestito
et al., \citealp{sestito} using open clusters). Most of the works seem to
converge on a value close to $\sim-0.06$~dex~kpc$^{-1}$ by using different
tracers (Cepheids by Luck et al. \citealp{luck}, Luck \& Lambert
\citealp{luck2011}, Yong et al. \citealp{yong}, Lemasle et
al. \citealp{lemasle} among others; open clusters by Friel et al.
\cite{friel}, Sestito et al. \citealp{sestito}, Pancino et al.
\citealp{pancino} among others; planetary nebulae by Pasquali \& Perinotto
\citealp{pasquali}, Maciel \& Quireza \citealp{maciel}; 
main-sequence turn-off stars by Cheng et al.
\citealp{cheng}, Co{\c s}kuno{\v g}lu et al. \citealp{coskunoglu}). 
It has been shown that the radial gradient becomes flatter with increasing distance
from the Galactic plane (Pasquali \& Perinotto \citealp{pasquali}; Boeche
\citealp{Boeche_PhD}; Cheng et al.  \citealp{cheng}) .  
If this flattening is caused by a zero radial gradient
in the thick disc (\citealp[Allende Prieto et al.][]{allende}; Ruchti et al.
\citealp{ruchti}; Bilir et al. 
\citealp{bilir}) or an effect of a flattening agent such as
radial mixing (as dynamical and chemo-dynamical models would suggest, see Sellwood \& Binney
\citealp{sellwood02}; Sch{\"o}nrich \& Binney \citealp{schoenrich}; 
S{\'a}nchez-Bl{\'a}zquez et al. \citealp{sanchez}; Minchev et
al. \citealp{minchev11a}, \citealp{minchev}) and energy feedback (Gibson et al.
\citealp{gibson}) is still unclear.\\
To explain the gradients observed in the Milky Way, 
an inside-out formation scenario of the Milky Way  has been proposed (Matteucci \& Fran\c{c}ois,
\citealp{matteucci}) in which the inner parts of the Galaxy experienced a
higher star formation rate than the outer parts.  Galactochemical models
(Chiappini et al.  \citealp{chiappini}, Cescutti et al., \citealp{cescutti})
and hydrodynamical simulations (Pilkington et al.
\citealp{pilkington}) with inside-out formation show
radial gradients consistent with the observed one.
The study by Sch\"onrich \& Binney \cite{schoenrich} claims that there is
no need to postulate an inside-out star formation to obtain negative
gradients. In their model the whole disc forms simultaneously,
and a metallicity gradient of $\sim-0.11$ dex kpc$^{-1}$ is finally
obtained because of the advection inwards of heavier elements
released by SNe.
Observational data (see Sestito et al. \citealp{sestito}, Luck
et al. \citealp{luck}, Carraro et al. \citealp{carraro}, 
Yong et al. \citealp{yong2012} among others) 
suggest that there are variations of the gradient along \R: 
a steeper gradient in the inner disc (\R$\lesssim5$~kpc) and a
flatter gradient in the outer disc (\R$\gtrsim10$~kpc).\\

In this paper we make use of the RAVE data to investigate the radial 
abundance gradients of the Milky Way.
RAVE is a large spectroscopic survey that collected
574\,630  spectra of 483\,330 stars of the Milky Way. The spectra are centred 
on the near-infrared \ion{Ca}{ii} triplet window (8410-8795\AA). The resolution of
R$\simeq$7500 allows measuring radial 
velocities (Steinmetz et al., \citealp{rave}), stellar parameters 
(Zwitter et al. \citealp{zwitter}; Siebert et al. \citealp{siebert}; Kordopatis et
al. \citealp{kordopatis13}), and chemical
abundances (Boeche et al. \citealp{boeche}; Kordopatis et
al. \citealp{kordopatis13}).
Distance estimates (Breddels et al. \citealp{breddels}; Zwitter et al.
\citealp{zwitter2010}; Burnett et al. \citealp{burnett}; Binney et al.
\citealp{binney2013}) enable one to locate the stars in three-dimensional space.
Proper motions are listed in a variety of
catalogues: Tycho2 (H\o g et al., \citealp{hog}), the 
PPM-Extended catalogues PPMX and PPMXL (Roeser et al., \citealp{roeser2008},
\citealp{roeser2010}), 
and the second and third U.S. Naval Observatory
CCD Astrograph Catalog UCAC2 and UCAC3 (Zacharias et al., \citealp{zacharias}).
Chemical  abundances for six elements are provided by the RAVE chemical pipeline
(Boeche et al. \citealp{boeche}; Kordopatis et al. \citealp{kordopatis13}), which
delivers abundances with errors of $\sim0.10-0.35$~dex, depending on the
element, stellar parameters, and signal-to-noise ratio (S/N).
The full set of kinematic and chemical data is a rich source of information
for studying the Galactic disc and has been recently employed for several
studies. By using the RAVE space and velocity data set, Siebert et al.
\cite{siebert2011} found a gradient in the mean Galactocentric radial velocity of stars in the
extended solar neighbourhood, which has been confirmed by a recent work of
Williams et al. \cite{williams} who, using red-clump giant stars, found that
it is more marked in the southern part of the disc, and furthermore
highlights wave-like
structures in the disc from the velocity distributions of the sample.
The structure of the disk was studied by Pasetto et al.
(\citealp{pasetto_a}, \citealp{pasetto_b}), Antoja et al. \cite{antoja},
Wilson et al. \cite{wilson}, and Veltz et al. \cite{veltz}.
The full set of RAVE chemo-kinematical data was recently used by
Boeche et al. \cite{boeche2013} to study the relation between abundances and
kinematics of the Galactic disc populations.
The radial metallicity gradient was previously studied in recent works by using
RAVE data (Karata{\c s} \& Klement \citealp{karatas}, Co{\c s}kuno{\v g}lu
et al. \citealp{coskunoglu}), but none of them investigated the gradients
for the individual elemental abundances.
Recently, Bilir et al. \cite{bilir} split
a sample of red-clump RAVE stars into thin- and thick-disc subsamples
by labelling the stars as a function of their kinematic characteristics,
and measured the metallicity gradients in these subsamples.\\

In the present work we investigate the chemical gradients of four elements
over the Galactocentric distance range of 4.5--9.5 kpc by using the chemical and kinematic
data of dwarf stars in the RAVE database. We also investigate
how the gradients change as function of the distance from the Galactic plane.
We applied the same analysis to the Geneva-Copenhagen 
Survey data set (Nordstr\"om et al. \citealp{nordstrom}) with the recent
data refinement by Casagrande et al. 
\cite{casagrande} and with a mock sample constructed with the code GALAXIA
(Sharma et al., \citealp{sharma}) to test the 
robustness of our results and compare them with Galactic
models.
We discuss how different stellar populations in the
sample can generate a bias and affect the gradients
measurements.
This paper will be followed by a second paper in which we study the
chemical gradients by using RAVE giant stars.\\

The paper is structured as follows: Section~2 describes the data and
the selection of the star samples used to measure the gradients,
in Section~3 we explain the analysis method applied to the samples,
in Section~4 and 5 we report the chemical gradients obtained,
in Section~6 we discuss biases and results and we conclude
in Section~7.

\begin{figure}[t]
\centering 
\resizebox{\hsize}{!}
{\includegraphics[clip]{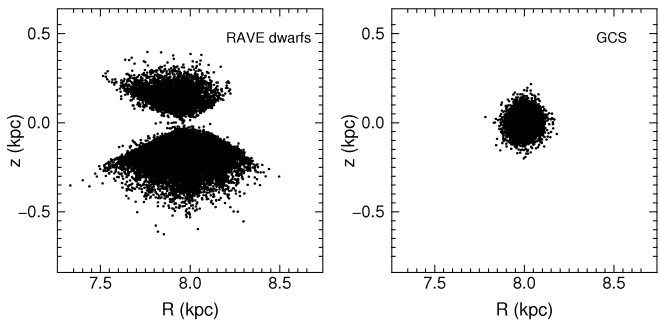}}
\caption{Spatial distribution of the selected 19\,962 RAVE stars (left) and
10\,616 stars of the Geneva-Copenhagen Survey (right) on the 
($R$,$z$).
}
\label{distr_xz_RAVE_casag} 
\end{figure}

\section{Data}

\begin{figure}[t]
\centering 
\resizebox{\hsize}{!}
{\includegraphics[clip]{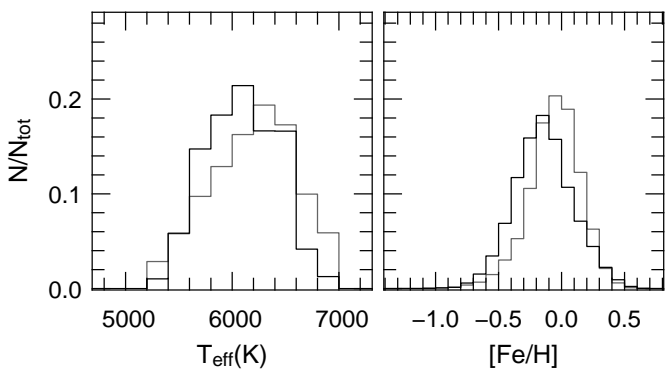}}
\caption{Temperature (left) and metallicity (right) 
distributions of the RAVE (black line) and GCS (grey line) samples.
}
\label{distrT_RAVE_casag} 
\end{figure}

\subsection{RAVE sample}
We selected the sample from the RAVE internal data archive, which contains
stellar atmospheric parameters (\temp, \logg\ and \met) of 451\,413 spectra,
and chemical elemental abundances for 450\,641 spectra.
The stellar atmospheric parameters are measured with the recently adopted
RAVE pipeline (Kordopatis et al. \citealp{kordopatis13}, and Kordopatis at al. 
\citealp{kordopatis13}). The chemical abundances ([X/H]) for Fe, Al, Mg, Si,
Ti, and Ni are derived by the RAVE chemical pipeline (Boeche et al. 
\citealp{boeche}, Kordopatis et al. \citealp{kordopatis13}).
We adopted the parallaxes by Binney et al. \citealp{binney2013} (submitted)
computed with an improved version of the recipe described in Burnett et al.
\cite{burnett}. \\
With these data we determined
Galactocentric positions and absolute velocities.
Galactic orbital parameters such as the guiding radius (\Rg) and maximum
distance from the Galactic plane (\Zmax) were considered and computed as explained in
Sec.~\ref{sec_error_orb}.
In this first paper we limit our investigation to dwarf stars with 
good-quality RAVE spectra (i.e. good stellar parameters and chemical abundances)
and reliable distances. The RAVE chemical catalogue provides abundances with
errors $\sigma\sim0.2$~dex for spectra with an
S/N$>$40. These errors hold for the elements Fe, Mg, Al, and Si (Ti and Ni 
measurements have low precision for dwarf stars).
Therefore, our selection criteria were 
i) stars with spectra with an S/N$>$40,
ii) dwarf stars with gravity \logg$>$3.8~dex and effective 
temperatures \temp\ from 5250 to 7000~K, and
iii) stars with distance uncertainties smaller than 30\%.
Moreover, we considered only the four elements Fe, Mg, Al, and Si because of
their good precision ($\sigma\sim0.15$~dex) within the selected \temp\ and S/N
ranges.
We took care that no peculiar stars were included in the sample by selecting
only spectra flagged as normal stars by Matijevi{\v c} et al.
\cite{matijevic}.\\
Among these stars, we furthermore selected only spectra for which the RAVE
pipeline converged to a single point of the
parameter space ($Algo\_Conv$ parameter equal to zero). These spectra are well fit
by the reconstructed spectra of the RAVE chemical pipeline ($\chi^2<2000$), and
with minor or no continuum defects\footnote{The $frac$ parameter   
described in Boeche et al. \cite{boeche} gives the fraction of pixels
that are non-defective and represents the goodness of the continuum
fitting.} ($frac>0.7$). Since RAVE re-observed $\sim$10\% of the objects,
some stars have two or more spectroscopic observations.
To avoid multiple observations in our sample, we chose only 
the spectrum with the highest S/N for re-observed stars. 
This selection yielded 19\,962 stars, whose 
spatial distribution is shown in Figure~\ref{distr_xz_RAVE_casag}. 
Most of the stars lie within 0.3 kpc from the Galactic plane 
and in the Galactic radius interval 
\R=7.6--8.3 kpc. This is a small interval where we can
estimate the local gradient, although its small spatial coverage
does not allow precise measurements. 
We extended the gradient measurements to the Galactic radius range
$R\sim$4.5-9.5 kpc by using the guiding radius \Rg.

\subsection{Geneva-Copenhagen Survey sample}
The GCS provides information such as the temperature,
metallicity, kinematics, distance, and age of 16\,682 F and G dwarf stars 
in the solar neighbourhood (Nordstr\"om et al., \citealp{nordstrom}).
We here refer to the more recent values reported by Casagrande et al.
\cite{casagrande}.
We selected stars by using the same constraints in
\temp, \logg, and guiding radii \Rg\ as were used for RAVE stars 
to keep the characteristics of the GCS and RAVE samples 
as similar as possible. With these constraints we selected
10\,616 stars. Still, there are some differences.
In fact, GCS stars are spherically distributed in space
and most of them do not lie farther away than 0.2~kpc
from the Galactic plane, whereas the RAVE stars have a
cone-shaped distribution in space and extend up to 0.5~kpc 
from the Galactic plane (Figure~\ref{distr_xz_RAVE_casag}).
Because most of the RAVE sample is inside the scale height of the thin disc
($\sim$0.3~kpc), it is reasonable to assume that the difference in stellar
populations between the two samples is small. We expect a slight
underrepresentation of the young, dynamically cold sub population in the RAVE
sample compared with the GCS sample because of the cone structure of the RAVE
volume.
In Figure~\ref{distrT_RAVE_casag} we compare the \temp\ and [Fe/H] distributions
of the two samples. The GCS sample has a temperature distribution
favouring higher \temp\ than the RAVE sample, and
the [Fe/H] distribution is shifted by $\sim$0.1~dex towards higher
metallicities. Despite the highlighted differences, the
GCS sample shows slightly flatter chemical gradients (but still
in reasonable agreement) than the RAVE sample, as we will
see below.

\subsection{RAVE mock sample}
The magnitude and spatial distributions of the RAVE sample may generate
observational biases that can lead to misinterpretations. To
avoid this, we created an equivalent mock sample with the
stellar population synthesis code GALAXIA \cite[Sharma et al.][]{sharma}, which uses
analytical density profiles based on the Besan\c{c}on model
\cite[Robin et al.][]{robin}. The mock sample reproduces the detailed RAVE selection
function in I magnitude and target density, both as functions of angular
position on the sky\footnote{The statistics on the celestial sphere were
derived with the HEALPix package \cite[G{\'o}rski et al.][]{gorski}.}.
Only RAVE spectra with S/N$>40$ and flagged as normal stars by
Matijevi{\v c} et al. \citep{matijevic} were considered.\\
The resulting mock catalog agrees well with the original RAVE
catalog for the stellar parameters and the distance and radial velocity
distributions. From this sample of 97\,485 entries, we imposed the same
selection criteria in \temp\ and \logg\ to mimic our
RAVE sample, obtaining a mock sample of 26\,198 entries. The higher
number of stars in the mock sample than the RAVE sample is due to
other selection criteria, that are applied to the RAVE sample (such as
$Algo\_Conv$ parameter convergence, $\chi^2$, $frac$ parameters and a
distance estimate) and not to the underlying Galaxy model.
The selected cut in S/N (S/N$>$40) helps minimising these differences, 
which should not affect the results on the abundance gradients significantly. 
In fact, although the spectrum quality parameters change as a function of
the distance (because RAVE is a magnitude-limited survey with constant
integration time, more distant stars have on average lower S/N spectra), this will
equally affect stars at any Galactocentric distance. However, to stay on the safe
side, we verified that 
the results were consistent with those obtained by applying the cut
when no cuts in distance error were applied to the RAVE data.

\section{Method and error estimation}

\subsection{Galactic orbit integration}\label{sec_error_orb} 
Because stars are oscillating radially and vertically along their orbits,
stellar samples in the solar neighbourhood contain information for a much
larger volume. To take advantage of this,
we integrated the Galactic orbits for a time of 40~Gyr (corresponding to
more than 50 periods for most of the stars) with the code NEMO
\cite[Teuben et al.][]{teuben}. As Galactic potential we adopted the
potential model n.2 by Dehnen \& Binney \cite{dehnen},
which assumes R$_\odot$=8.0 kpc, 
and best matches the observed properties of the Galaxy. We used the same
potential for all samples, including the mock sample based on the Besancon
model. In axisymmetric equilibrium models the z-component of the angular momentum
$L\mathrm{z}$ and the vertical action $J\mathrm{z}$ are constants of motion.
With the help of the rotation curve $v_\mathrm{c}(R_\mathrm{g})$ the angular
momentum can be converted into the guiding radius \Rg. The commonly used mean radius
$R_\mathrm{m}$ of apocentre $R_\mathrm{a}$ and pericentre $R_\mathrm{p}$ is
a proxy for the guiding radius only for low eccentricities, therefore \Rg\
has to be preferred. With
these quantities we can study the gradients at Galactic radii
beyond the range currently occupied by the stars.
Vertically, the longest distance to the Galactic plane \Zmax\ is
a measure of the vertical action.\\
The values \Rg\ and \Zmax\ were computed with the method 
just described for the RAVE, the GCS, and the mock
samples.
\\

\subsection{Gradient error estimation}\label{sec_error_boot} 
We measured the
abundance gradients by fitting the distribution of the stars in the
(\Rg,[X/H]) plane with a linear regression.  We also investigated more
sophisticated methods such as modelling the density distribution of the stars
with a two-dimensional function such as a Gaussian or lognormal function, and
performing the fitting procedure with the maximum-likelihood method.  The
different methods provided consistent results. The linear regression was
ultimately preferred because it is simpler and faster, in particular during the
confidence interval evaluation. The gradient confidence intervals reported
in this paper were evaluated by using a
resampling method such as bootstrapping.  Means and 1$\sigma$ errors 
(68\% confidence) were computed over 1000 realizations 
for every gradient reported. These errors represent the internal
errors, as explained in the following section.

\subsection{Bias in the gradients and gradient errors}\label{sec_err_bias}
One needs to be aware that the use of the guiding radius 
\Rg\ introduces a kinematic bias for volume-limited samples such as the RAVE
and GCS samples. Stars with small \Rg\ can
reach the solar neighbourhood only if they move in eccentric
(kinematically hot) orbits. Such stars are on average more metal-poor than
stars at the same \Rg\ and in circular orbits, which will never
reach the solar neighbourhood. This causes a lack of metal-rich stars
at small \Rg\ in volume-limited samples, and for these samples the measured
gradient is less negative.
The lack of metal-rich stars occurs at large \Rg\ as well, 
with the difference that in the outer Galaxy the stellar density
is lower, therefore the effect is weaker.
Although we cannot precisely quantify this bias, we discuss 
the impact of this bias on the measured gradients.

\section{Analysis and results}\label{sec_analysis}
In this section we split the samples into three ranges of
\Zmax, discuss the [Fe/H] distributions, and derive the radial gradients in
[Fe/H] and in [$\alpha$/Fe].
In Sec.~\ref{sec_boxcar} we refine the method for the RAVE sample to improve the
vertical resolution and extend the analysis to the elements Fe, Al,
Mg, and Si and the relative abundance gradients [X/Fe].

\subsection{Gradients from the RAVE sample}
\begin{figure*}[t]
\centering
\includegraphics[width=12cm,clip]{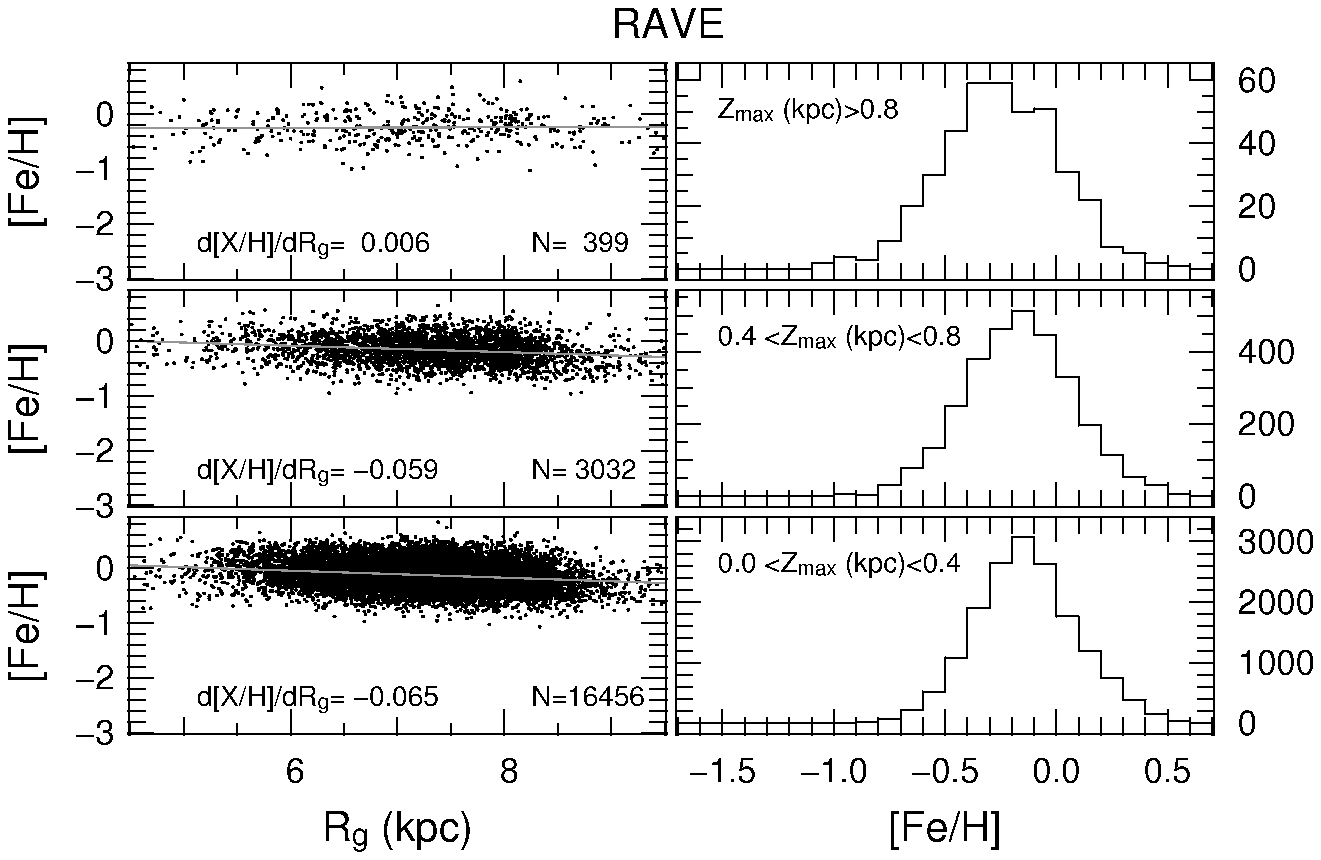}
\includegraphics[width=12cm,clip]{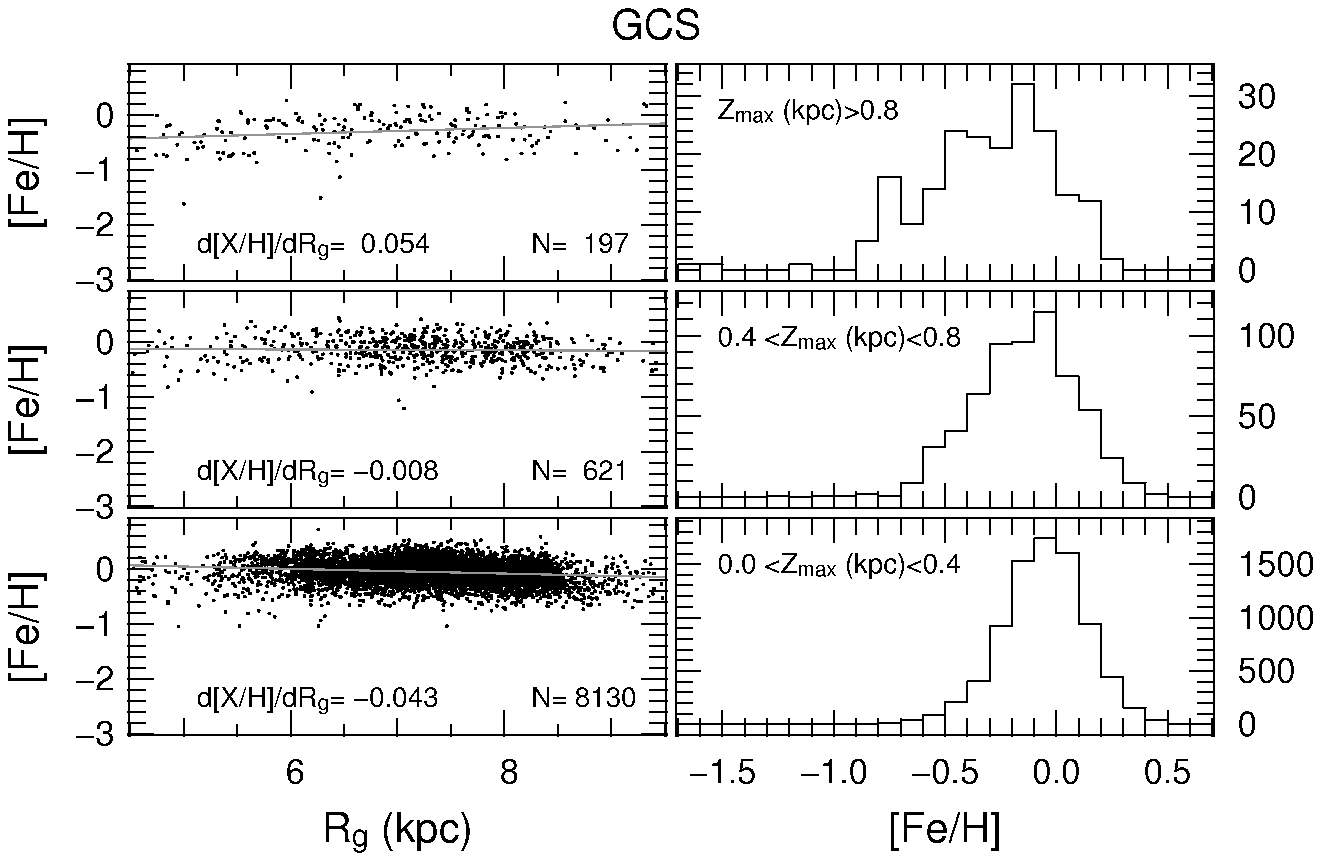}
\includegraphics[width=12cm,clip]{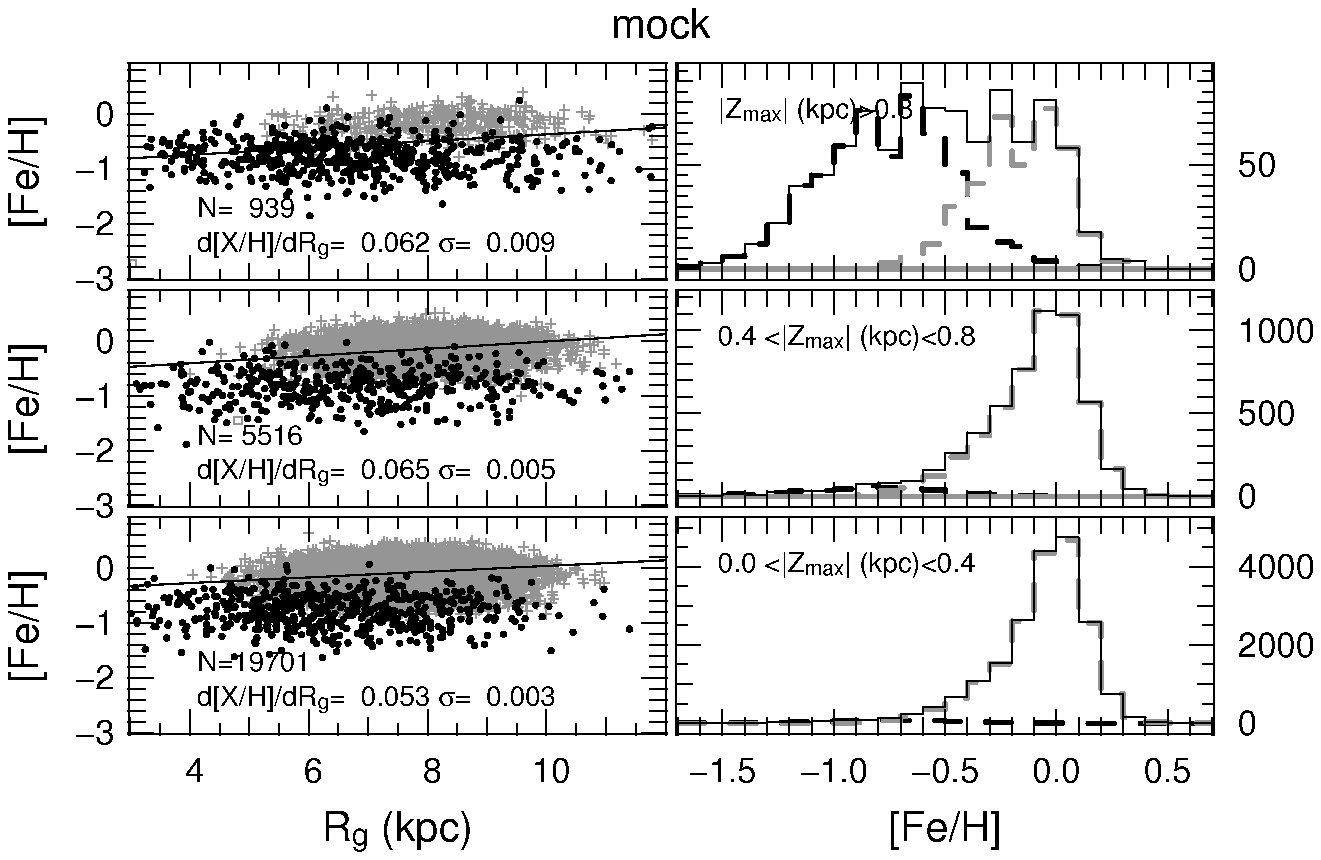}
\caption{Distributions of the dwarf stars for the RAVE sample (top), GCS
sample (middle) and mock sample (bottom) in the ([Fe/H],\Rg) planes (left
panels) and [Fe/H] distributions (right panels). The grey line has a slope
equal to the gradient. For the mock sample (bottom) grey plus symbols, 
black circles, and grey squares represent thin-disc, thick-disc, and halo stars, respectively.
The same order is followed by the grey dashed, black dashed, and grey solid
lines in the right panels, which represent the iron distributions of these
subsamples.}

\label{fitting} 
\end{figure*}

The resulting gradient for iron at the present position of the stars is
$\frac{d [Fe/H]}{d R}= -0.029\pm0.011$ dex kpc$^{-1}$. The small
range in \R\ covered by the sample (7.6$<$\R(kpc)$<$8.3) does not allow a precise measurement.
The derived metallicity gradient is significantly flatter than
expected, but
this may be due to the conical volume, where a negative vertical gradient
biases the radial gradient to lower values.
Moreover, \R\ may not be representative of the distance at which the
stars were born, since many of them do not have circular orbits.
From the probability point of view, the guiding radius \Rg\ better 
represents the Galactocentric distance of the
star during one revolution around the Galactic centre.
By using \Rg\ and the constraints 4.5$<R_\mathrm{g}$(kpc)$<$9.5
(to avoid the few outlier stars that can affect the estimation)
we obtained the result\\
$\frac{d [Fe/H]}{d R_\mathrm{g}}= -0.059\pm0.002$ dex kpc$^{-1}$.\\
We found in a previous work (Boeche, \citealp{Boeche_PhD}) 
that different gradients are obtained at different \Zmax.
We divided the full sample into three subsamples with different \Zmax ranges:
$0.0<Z_\mathrm{max}$ (kpc)$\leq 0.4$ (16\,456 stars, named $Z_\mathrm{0.0}^\mathrm{RAVE}$ sample) 
$0.4<Z_\mathrm{max}$ (kpc)$\leq 0.8$ (3032 stars, the $Z_\mathrm{0.4}^\mathrm{RAVE}$
sample),
and $Z_\mathrm{max}$ (kpc)$>0.8$ (399 stars, the $Z_\mathrm{0.8}^\mathrm{RAVE}$
sample).
For these samples we found\\
$\frac{d [Fe/H]}{d R_\mathrm{g}}
\mbox{($Z_\mathrm{0.0}^\mathrm{RAVE}$)}=-0.065\pm0.002$ dex kpc$^{-1}$\\
$\frac{d [Fe/H]}{d R_\mathrm{g}}\mbox{($Z_\mathrm{0.4}^\mathrm{RAVE}$)}=-0.059\pm0.005$ dex kpc$^{-1}$\\
$\frac{d [Fe/H]}{d R_\mathrm{g}}\mbox{($Z_\mathrm{0.8}^\mathrm{RAVE}$)}=0.005\pm0.012$ dex
kpc$^{-1}$.\\
In Figure~\ref{fitting} the gradients are represented 
by the slope of the straight line (left panels).
We also report the gradients for [$\alpha$/Fe]\footnote{We recall that in
RAVE the $\alpha$-enhancement is computed as
[$\alpha$/Fe]=$\frac{[Mg/Fe]+[Si/Fe]}{2}$.}:
\\
\noindent
$\frac{d[\alpha/Fe]}{dR_\mathrm{g}}\mbox{($Z_\mathrm{0.0}^\mathrm{RAVE}$)}
=-0.004\pm0.001$ dex kpc$^{-1}$,\\
$\frac{d[\alpha/Fe]}{dR_\mathrm{g}}\mbox{($Z_\mathrm{0.4}^\mathrm{RAVE}$)} =
-0.005\pm0.002$ dex kpc$^{-1}$,\\
$\frac{d[\alpha/Fe]}{dR_\mathrm{g}}\mbox{($Z_\mathrm{0.8}^\mathrm{RAVE}$)} =
-0.020\pm0.005$ dex kpc$^{-1}$.\\

These results confirm that the radial gradients in the RAVE sample vary with
increasing $Z_\mathrm{max}$. The complete set of measured gradients
of abundances [X/H] and relative abundance [X/Fe] for the RAVE elements are
given in Tables~\ref{tab_XH_grad_Rg} and \ref{tab_XFe_grad_Rg}.\\

\subsection{Gradients from the Geneva-Copenhagen survey sample}
To evaluate the robustness of our results we repeated the
same procedure for the GCS data.  
With the actual Galactocentric radius \R\ of all stars,
the gradient is $\frac{d [Fe/H]}{d R}=-0.004\pm0.047$ dex kpc$^{-1}$,
which is unreliable because the very small radial extent of the sample.\\
With the guiding radius in the range \Rg$\sim$4.5-9.5 kpc we obtained
a gradient
$\frac{d [Fe/H]}{d R_\mathrm{g}}=-0.027\pm0.004$ dex kpc$^{-1}$ .\\
We followed the procedure used for the RAVE sample and selected three
subsamples in the following \Zmax\ ranges:
$0.0<Z_\mathrm{max}$ (kpc)$<0.4$ (8130 stars, named the $Z_\mathrm{0.0}^\mathrm{GCS}$ sample) 
$0.4<Z_\mathrm{max}$ (kpc)$<0.8$ (621 stars, the $Z_\mathrm{0.4}^\mathrm{GCS}$ sample)
and $Z_\mathrm{max}$ (kpc)$>0.8$ (197 stars, the $Z_\mathrm{0.8}^\mathrm{GCS}$ sample). 
The resulting gradients are (see also Figure~\ref{fitting}, middle)\\
$\frac{d[Fe/H]}{dR_\mathrm{g}}\mbox{($Z_\mathrm{0.0}^\mathrm{GCS}$)} =-0.043\pm0.004$ dex
kpc$^{-1}$,\\
$\frac{d[Fe/H]}{dR_\mathrm{g}}\mbox{($Z_\mathrm{0.4}^\mathrm{GCS}$)} =
-0.008\pm0.011$ dex kpc$^{-1}$,\\
$\frac{d[Fe/H]}{dR_\mathrm{g}}\mbox{($Z_\mathrm{0.8}^\mathrm{GCS}$)} =
+0.056\pm0.019$ dex kpc$^{-1}$.\\
The metallicity distribution shows a more densely populated low-metallicity tail at
large  $Z_\mathrm{max}$ than the RAVE sample.
The gradients $d[Fe/H]/dR_\mathrm{g}$ are in general less negative
than the RAVE sample. The trend of a flatter gradient with
$Z_\mathrm{max}$ agrees with the RAVE data, with the exception of
the $Z_\mathrm{0.8}^\mathrm{GCS}$ subsample (which has a positive gradient), 
which is particularly
affected by the bias discussed in Sec.~\ref{sec_err_bias}.
This gradient is probably not reliable,
and in Sec.~\ref{sec_obs} we provide a more realistic
value.\\

We tentatively measured the gradients of the $\alpha$-elements
given by Casagrande et al. \cite{casagrande}, 
with the warning that these abundances were estimated from photometry and
must be considered only a proxy of [$\alpha$/Fe]. 
For [$\alpha$/Fe] we obtained\\ 
\noindent
$\frac{d[\alpha/Fe]}{dR_\mathrm{g}}\mbox{($Z_\mathrm{0.0}^\mathrm{GCS}$)}
=+0.010\pm0.002$ dex kpc$^{-1}$,\\
$\frac{d[\alpha/Fe]}{dR_\mathrm{g}}\mbox{($Z_\mathrm{0.4}^\mathrm{GCS}$)} =
-0.006\pm0.005$ dex kpc$^{-1}$,\\
$\frac{d[\alpha/Fe]}{dR_\mathrm{g}}\mbox{($Z_\mathrm{0.8}^\mathrm{GCS}$)} =
-0.023\pm0.007$ dex kpc$^{-1}$.\\

Using only stars with more reliable temperature estimates (treated with
the infrared flux method of Casagrande et al.  \citealp{casagrande10}), the
results do not change significantly (differences smaller than 1$\sigma$).\\

\subsection{Gradients from the RAVE mock sample}
The mock sample allows us to separate the influence of the different
components (thin disc, thick disc, stellar halo) on the properties of the
RAVE sample. In Figure~\ref{fitting} (bottom panel) the contribution 
of the three components are separately shown in every \Zmax\ slice.
The halo contribution is
negligible, and the thick disc is detected as a prominent low-metallicity peak
at \Zmax$>0.8$ kpc. The Besan\c con model assumes a radial
metallicity gradient of d[Fe/H]/dR= --0.07 dex kpc$^{-1}$ for the thin disc and no
gradient in the thick disc. The gradients we found deviate significantly
from these values for reasons that we analyse in Sec.~\ref{sec_mock}.
Using the actual galactocentric radius \R,
the gradient for the mock sample is 
$\frac{d [Fe/H]}{d R}=-0.156\pm0.010$ dex kpc$^{-1}$, significantly more
negative than expected. This is probably caused by the
uneven spatial coverage in RAVE.\\
Using the guiding radius \Rg, we obtained a gradient
$\frac{d [Fe/H]}{d R_\mathrm{g}}=+0.060\pm0.003$ dex kpc$^{-1}$ .\\
Selecting three subsamples in the three different \Zmax\ ranges
as we did before, the resulting gradients are\\
$\frac{d[Fe/H]}{dR_\mathrm{g}}\mbox{($Z_\mathrm{0.0}^\mathrm{mock}$)}
=+0.053\pm0.003$ dex
kpc$^{-1}$,\\
$\frac{d[Fe/H]}{dR_\mathrm{g}}\mbox{($Z_\mathrm{0.4}^\mathrm{mock}$)} =
+0.065\pm0.005$ dex kpc$^{-1}$,\\
$\frac{d[Fe/H]}{dR_\mathrm{g}}\mbox{($Z_\mathrm{0.8}^\mathrm{mock}$)} =
+0.063\pm0.008$ dex kpc$^{-1}$.\\

We were unable measure the [$\alpha$/Fe] gradient for the mock sample 
because the Besan\c con model does not provide this observable.\\

\begin{table*}[t]
\caption[]{Radial abundance gradients measured in the RAVE sample for Fe, Mg, Al, and Si 
expressed as dex kpc$^{-1}$ for three
ranges of \Zmax. Uncertainties of 68\% confidence are
obtained with the bootstrap method. 
}
\label{tab_XH_grad_Rg}
\vskip 0.3cm
\centering
\begin{tabular}{l|ccccc}
\hline
\noalign{\smallskip}
        &  $\frac{d[Fe/H]}{dR_\mathrm{g}}$  & $\frac{d[Mg/H]}{dR_\mathrm{g}}$  & $\frac{d[Al/H]}{dR_\mathrm{g}}$
& $\frac{d[Si/H]}{dR_\mathrm{g}}$ \\
\noalign{\smallskip}
\hline
\noalign{\smallskip}
0.0$\leq Z_\mathrm{max}\mathrm{(kpc)}<$0.4           &$-0.065\pm0.003$&$-0.073\pm0.002$&$-0.085\pm0.003$&$-0.064\pm0.003$\\

0.4$\leq Z_\mathrm{max}\mathrm{(kpc)}<$0.8           &$-0.059\pm0.006$ &$-0.061\pm0.005$&$-0.076\pm0.007$ &$-0.063\pm0.005$\\

\phantom{0.4$\leq$}$Z_\mathrm{max}\mathrm{(kpc)}>$0.8& $+0.006\pm0.015$&$-0.026\pm0.011$&$-0.017\pm0.016$ & $-0.008\pm0.013$\\
\noalign{\smallskip}
\hline
\end{tabular}

\end{table*}

\begin{table*}[t]
\caption[]{As Table~\ref{tab_XH_grad_Rg}, but for relative abundances [X/Fe].}
\label{tab_XFe_grad_Rg}
\vskip 0.3cm
\centering
\begin{tabular}{l|ccccc}
\hline
\noalign{\smallskip}
        &  & $\frac{d[Mg/Fe]}{dR_\mathrm{g}}$ & 
$\frac{d[Al/Fe]}{dR_\mathrm{g}}$ & $\frac{d[Si/Fe]}{dR_\mathrm{g}}$ \\
\noalign{\smallskip}
\hline
\noalign{\smallskip}
0.0$\leq Z_\mathrm{max}\mathrm{(kpc)}<$0.4            &       &$-0.009\pm0.002$&$-0.022\pm0.002$&$+0.001\pm0.001$\\

0.4$\leq Z_\mathrm{max}\mathrm{(kpc)} <$0.8            &       & $-0.004\pm0.003$&$-0.018\pm0.004$ & $-0.004\pm0.003$\\

\phantom{0.4$\leq$}$ Z_\mathrm{max}\mathrm{(kpc)} >$0.8 &       &$-0.027\pm0.007$&$-0.024\pm0.009$ &$-0.013\pm0.006$\\
\noalign{\smallskip}
\hline
\end{tabular}

\end{table*}

\section{Gradient estimates with moving box car}\label{sec_boxcar}
\begin{figure*}[t]
\begin{minipage}[t]{8cm}
\includegraphics[width=8cm,clip]{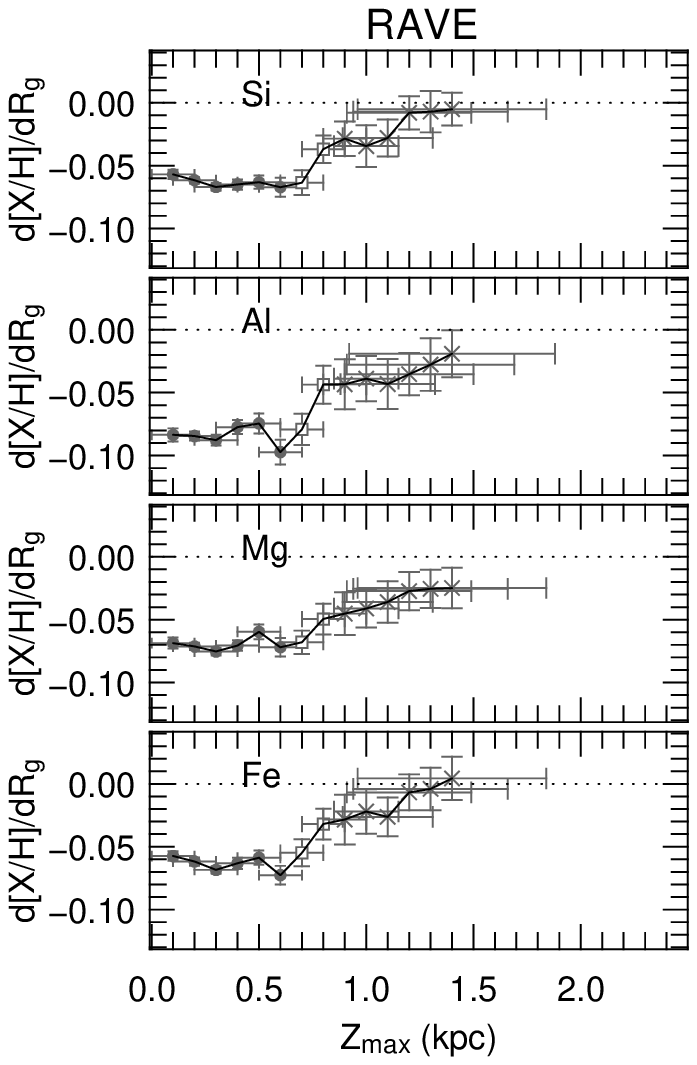}
\includegraphics[width=8cm,clip]{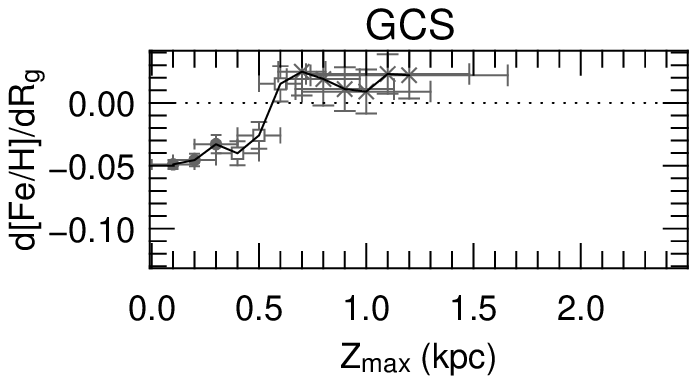}
\includegraphics[width=8cm,clip]{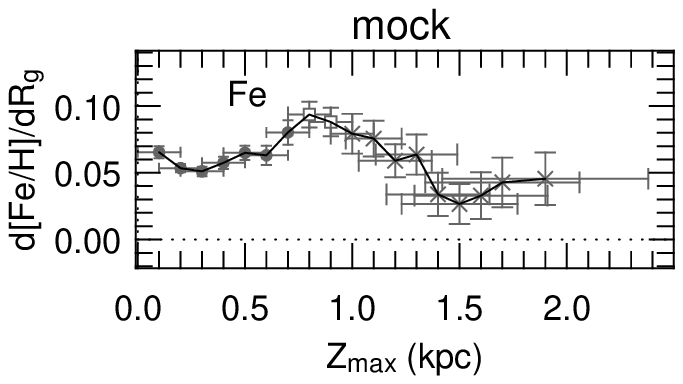}
\caption{Abundance gradients for Fe, Mg, Al and Si for the RAVE sample
(top), [Fe/H] for GCS sample (middle), and for the mock sample
(bottom) as a function of \Zmax. Horizontal error bars indicate the \Zmax\
interval in which the gradient has been measured. Vertical error bars indicate the
gradient error estimated with the bootstrap technique outlined in
Sec.\ref{sec_error_boot}. Full points indicate that the gradient has been
computed with a sample containing N$>$1000 stars; open square represent
samples with 1000$<$N$<$300 stars; cross symbols represent samples with 200$<$N$<$300
stars.}
\label{Zmax_FeH_grad_bcar}
\end{minipage}
\hfill
\begin{minipage}[t]{8cm}
\includegraphics[width=8cm,clip]{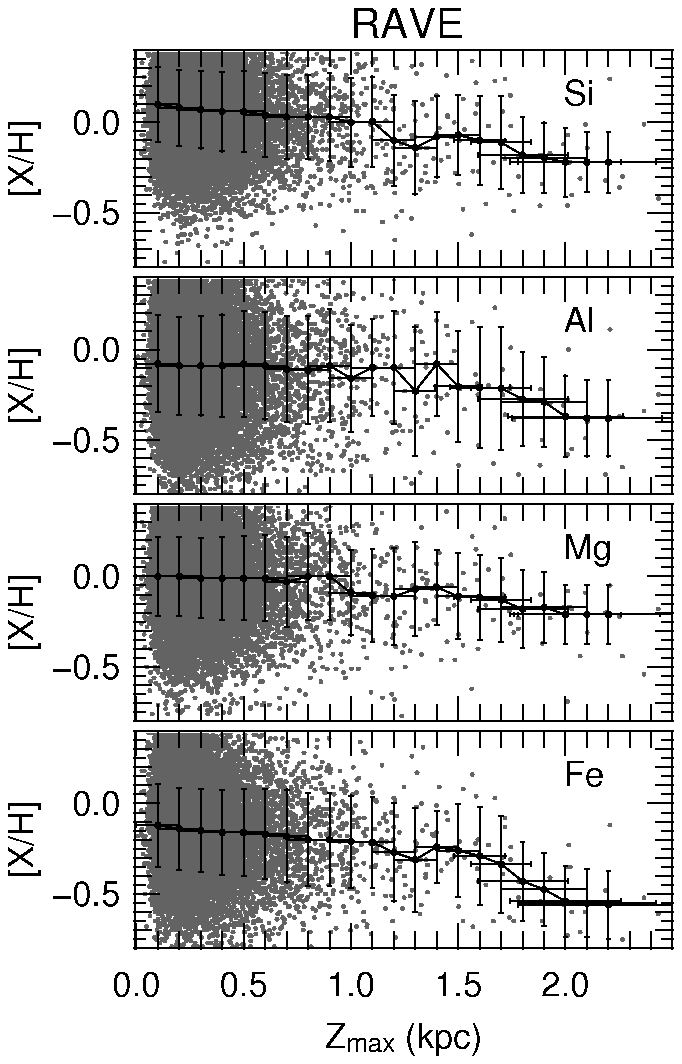}
\includegraphics[width=8cm,clip]{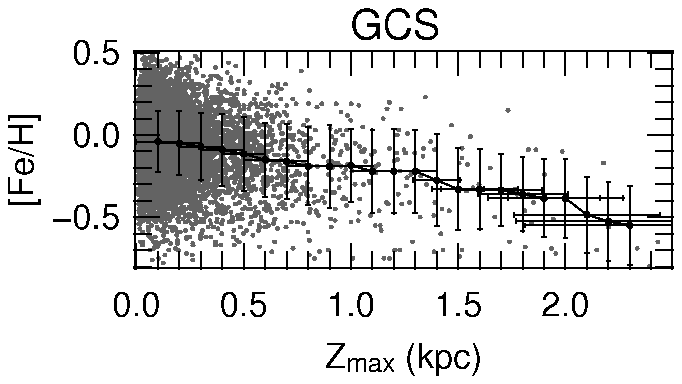}
\includegraphics[width=8cm,clip]{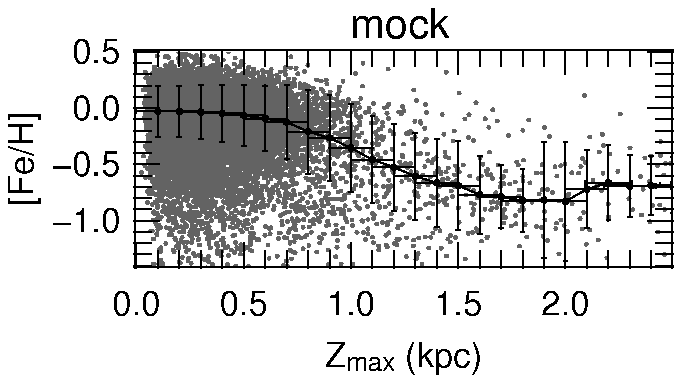}
\caption{Abundance for the elements Fe, Mg, Al, and Si for the RAVE
sample (top), [Fe/H] for the GCS sample (middle) and the mock sample
(bottom) as a
function of \Zmax. The black dots connected by a line represent the median
abundance. The vertical error bars indicate the standard deviation
of the distribution computed in the \Zmax\ interval indicated by the
horizontal error bars.}
\label{Zmax_XH_bcar}
\end{minipage}
\end{figure*}

The previous results show that the gradient flattens when
moving to higher \Zmax. Whether this change occurs smoothly or with
a sudden transition can be investigated by measuring the gradient in an
interval of constant width, centred on progressively larger distances from
the plane. \\
We measured the gradient $\frac{d[X/H]}{dR_\mathrm{g}}$ of a sample of stars lying
in the interval [\Zmax-0.1,\Zmax+0.1] and moved this sampling
interval by 0.1~kpc steps from 0.1~kpc towards larger \Zmax. 
To always have a statistically
significant sample in these intervals, we imposed the condition that at every
step the sample must contain no fewer than 200 stars. If there were fewer stars,
the \Zmax\ interval was extended until the condition was matched.
Undersampling occurs in particular at \Zmax$>$0.7~kpc, where the number of
stars is small. We present the result in Figure~\ref{Zmax_FeH_grad_bcar}, 
where the gradient
$\frac{d[X/H]}{dR_\mathrm{g}}$ is shown as a function of \Zmax.
We found that the gradients remained constant for all elements until
\Zmax$\sim 0.7$~kpc,
then they increase to shallower (or positive) gradients with a hint of
a step occurring between 0.6 and 0.9 kpc. 
For [Fe/H] the gradient of the RAVE sample is --0.06 dex kpc$^{-1}$
and $\sim0$ at small and large \Zmax, respectively, slightly more
negative than the GCS sample, where it amounts to --0.05 and +0.02 dex
kpc$^{-1}$, respectively.
The mock sample shows positive gradients at any \Zmax. The value of +0.06
dex kpc$^{-1}$ at low \Zmax\ decreases to +0.02 at large \Zmax, where the
thick disc dominates.\\
In Figure~\ref{Zmax_XH_bcar} we show the abundance distributions [X/H]
as a function of \Zmax\ for the four elements of the RAVE sample and for [Fe/H] in the
GCS and the mock sample. In this figure we allowed a minimum 
of 20 stars in the interval to see local changes in the mean abundance.
There is a clear negative trend visible in all
cases. However, the trend of [Fe/H] for the GCS sample is slightly smoother
but covers the same range. The mock sample behaves quite differently: 
although the mean iron abundance
at small and large \Zmax\ is similar to that of the RAVE and GCS samples
($\sim0.6$~dex difference between \Zmax=0 and \Zmax=2~kpc), 
its evolution as a function of \Zmax\ is quite different.
While for RAVE and GCS the decrease of the mean Fe abundance in the range 
0.0$<$\Zmax$<$1.6 kpc appears to be approximately linear, 
for the mock sample it has a negative
second derivative up to \Zmax$\sim$0.6 kpc and becomes positive afterwards. 
This difference suggests that the transition between the
thin and thick disc in RAVE and GCS differs from the prediction of the
mock sample.\\

\section{Discussion}\label{grad_result}
An important goal of investigating the chemical gradients
is to distinguish the effect of biases caused by the selection
function, the impact of the thin
disc/thick disc dichotomy, and the features of the intrinsic evolution of
the thin disc by an inside-out growth (Matteucci \& Fran\c{c}ois,   
\citealp{matteucci}).
The three samples analysed here show some similarities, but
also striking differences.

\subsection{Abundance gradients in the GALAXIA/Besan\c con mock sample}\label{sec_mock}
In the local sample of any Milky Way model the spatial structure of
components or sub populations are represented only indirectly and only if the
model is set up in a dynamically consistent way. The radial scalelength of
the density in the disc is represented by a proper choice of the asymmetric
drift (the shift of the tangential velocity distribution with respect to the
circular speed). Vertically, the thickness corresponds to the assigned
velocity dispersion.
In addition to the density distributions and
kinematics, stellar metallicities are assigned to each sub population
according to their current position, in the Besan\c con model, regardless of
kinematics.
The vertical metallicity distribution is set up in a consistent
way, because each sub population is assigned a spatially homogeneous metallicity
distribution. Radially, an uncorrelated local metallicity distribution would
also require a homogeneous metallicity distribution. The additionally
assigned spatial metallicity gradient of d[Fe/H]/dR=--0.07 dex kpc$^{-1}$ 
for the thin-disc sub populations would require the
correlation of the metallicity with the tangential velocity or angular
momentum to be dynamically consistent. This means that the local sample with no
correlation of tangential velocity and metallicity looks like a sample of
components with no radial metallicity gradient.
This caveat could easily be overcome by assigning the metallicity distribution
at each location as a function of the angular momentum (or tangential
velocity), leading to a radial gradient with respect to \Rg. The
radial gradient as a function of the current position $R$ would then be very
similar (see e.g. Binney \citealp{binney2010} for a dynamical model).

Nevertheless, we can learn about the impact of a distinct thick disc with
lower metallicity and higher velocity dispersion and stronger asymmetric drift
on the mock sample. 
The thick disc with a mean metallicity of --0.78~dex and 0.3~dex internal dispersion
has an asymmetric drift of 53~km~s$^{-1}$ and a tangential velocity dispersion
of 51~km~s$^{-1}$ compared with 15 and 28~km~s$^{-1}$ for the old thin disc.
This shifts the thick-disc stars to the lower left in the (\Rg,[Fe/H])
diagrams (Figure~\ref{fitting} bottom) with respect to the thin-disc stars,
generating an apparent decreasing mean [Fe/H] with
decreasing \Rg, or in other words, a fictitious positive gradient. 
This trend is stronger for larger \Zmax,
because the fraction of thick-disc stars increases with a still dominating
thin disc at large \Rg.
For the analysis of thin disc properties the contamination by thick disc
stars in the RAVE sample results in a bias towards less negative or even positive
radial metallicity gradients with respect to \Rg.

An additional striking difference of the mock sample compared with the 
RAVE and the GCS samples is the large number of low-metallicity stars for
\Zmax$>$0.8~kpc (see Figure~\ref{fitting}).
The reasons for this deviation can be the i) thick-disc density, ii)
its vertical velocity dispersion, and iii) its mean metallicity adopted by the 
Besan\c con model. A lower local density in the model
(at present assumed to be $\sim$10\%) would reduce the peak proportionally, a
lower
vertical velocity dispersion (at present assumed to be 42~km~s$^{-1}$) would shift a
significant part to the lower \Zmax\ bin and slightly lift the 
low-metallicity tail, whereas a higher mean metallicity 
(at present assumed --0.78dex) would merge the peak with the thin-disc distribution.

Previous authors have already shown that the thick
disc might have a higher mean metallicity than the one assumed in the Besan\c con
model. A thick-disc mean metallicity of $\sim-0.5$~dex
was found by some authors (\citealp[Bilir et al.][]{bilir}; 
\citealp[Kordopatis et al.][]{kordopatis_b}; \citealp[Schlesinger et
al.][]{schlesinger}; \citealp[Soubiran et al.][]{soubiran03}) with a gradient
that decreases the mean metallicity $\sim-0.8$~dex beyond 3~kpc (\citealp[Katz et
al.][]{katz}; \citealp[Ivezi{\'c} et al.][]{ivezic}). The lack of
a prominent low-metallicity tail in the RAVE and GCS samples compared with
the mock sample supports these studies.  

Despite the unrealistic gradients exhibited,
the mock sample gives very precious information for data interpretation. 
It suggests that short scale-length metal-poor components (thick disc
and halo stars) may accumulate at low \Rg, and, together with the thin-disc
stars, cause a
fictitious flatter or positive gradient that depends on the relative fraction of 
the stellar populations in the sample. This might be one of the possible
explanations of the positive
metallicity gradients in thick-disc-like stars found in some previous works
(Nordstr\"om et al. \citealp{nordstrom}; Carrell et al. \citealp{carrell}). 
In observational data the
population mix depends on the selection function of the sample under
analysis, and only a complete (ideal) sample can deliver unbiased 
gradients. We must therefore be careful when interpreting the gradient
measurement of any sample available today.\\

\subsection{Abundance gradients of the observational
samples}\label{sec_obs}
In the previous section we saw how the mix of thin- and thick-disc stars
in the Besan\c con model causes a positive gradient when
the kinematically dependent quantity \Rg\ is used. 
This effect is enhanced in our samples by the observational bias discussed
in Sec.~\ref{sec_err_bias}: metal-rich stars in (nearly) circular orbits at small \Rg\ are missing
because they cannot reach our volume-limited samples located in the solar
neighbourhood, and the gradient becomes flatter than it should be.
This bias is clearly visible in the GSC sample (Figure~\ref{fitting}, middle
panel)
where stars at 4.5$\leq$\Rg(kpc)$\leq$5.5 are clearly [Fe/H] poorer 
than stars at higher \Rg. 
Excluding this \Rg\ interval from the measurement, the
gradients for the GCS data are
$d [Fe/H]/d R_\mathrm{g}=-0.054\pm0.003$ dex kpc$^{-1}$,      
$-0.028\pm0.012$ dex kpc$^{-1}$, and $+0.004\pm0.025$ dex kpc$^{-1}$ for the
subsamples $Z_\mathrm{0.0}^\mathrm{GCS}$, $Z_\mathrm{0.4}^\mathrm{GCS}$, and    
$Z_\mathrm{0.8}^\mathrm{GCS}$, respectively, whereas for the RAVE data
they are $d [Fe/H]/d R_\mathrm{g}=-0.066\pm0.003$ dex kpc$^{-1}$,
$-0.063\pm0.006$ dex kpc$^{-1}$, and $-0.002\pm0.016$ dex kpc$^{-1}$ for the
subsamples $Z_\mathrm{0.0}^\mathrm{RAVE}$, $Z_\mathrm{0.4}^\mathrm{RAVE}$, and
$Z_\mathrm{0.8}^\mathrm{RAVE}$, respectively.
The new GCS results are closer to those found with the RAVE data
and, in particular, the peculiar positive gradient found with the
$Z_\mathrm{0.8}^\mathrm{GCS}$ sample disappears. We verified that
adopting  other cuts at low or high \Rg\ causes 
changes in gradient of the order of 2-3~$\sigma$ for both the RAVE and the GCS
samples (that means $\sim0.01$~dex kpc$^{-1}$).
These results suggest that the errors obtained with the bootstrap method
(explained in Sec.~\ref{sec_error_boot}) represent the internal
errors, in which observational biases are neglected.
Local inhomogeneities in the stellar population in the solar neighbourhood 
(such as disrupted open clusters and  moving groups of kinematic origin, see Famaey et al. 
\citealp{famaey}; Antoja et al. \citealp{antoja}) can also explain the differences 
in gradient observed between the RAVE and the GCS
samples and between different \Rg\ intervals considered. 
Our conclusion is that these observational biases can affect the measured gradients
for a quantity of about $\sim0.01$~dex kpc$^{-1}$, which is larger
than our internal errors, but it does not affect the interpretation of
our results.

\subsection{Radial chemical gradients of the RAVE
sample as a function of \Zmax}
\begin{figure*}[t]
\begin{minipage}[t]{8cm}
\includegraphics[width=8cm,clip]{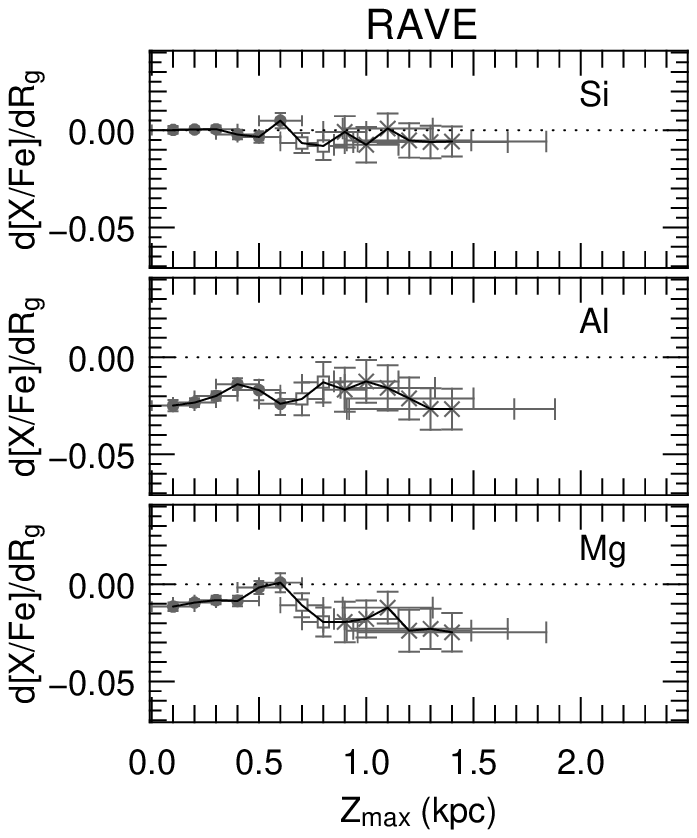}
\includegraphics[width=8cm,clip]{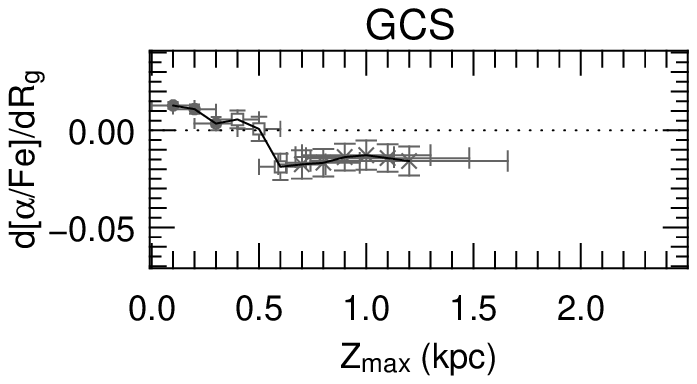}
\caption{Gradients for the RAVE relative abundances [Mg/Fe], [Al/Fe], and
[Si/Fe] for RAVE (top), and [$\alpha$/Fe] for GCS (bottom) as a function of \Zmax. 
Horizontal and vertical error bars are defined as in Figure~\ref{Zmax_FeH_grad_bcar}.}
\label{Zmax_XFe_grad_bcar}
\end{minipage}
\hfill
\begin{minipage}[t]{8cm}
\includegraphics[width=8cm,clip=]{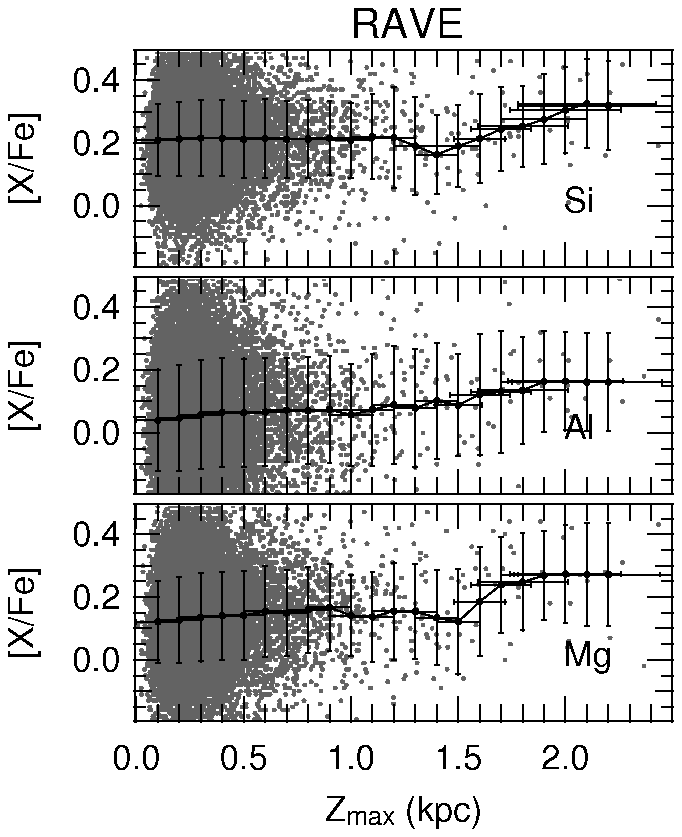}
\includegraphics[width=8cm,clip]{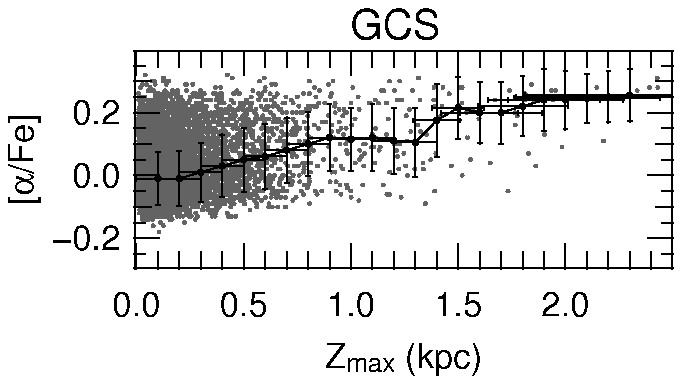}
\caption{Relative abundances [Mg/Fe], [Al/Fe], and [Si/Fe] for RAVE (top),
and [$\alpha$/Fe] for GCS (bottom) as a function of \Zmax.
Horizontal and vertical error bars are defined as in
Figure~\ref{Zmax_XH_bcar}.}
\label{Zmax_XFe_bcar} 
\end{minipage}
\end{figure*}

In Section~\ref{sec_analysis} we began the analysis by dividing 
the sample into three subsamples
$Z_\mathrm{0.0}^\mathrm{RAVE}$, $Z_\mathrm{0.4}^\mathrm{RAVE}$, and
$Z_\mathrm{0.8}^\mathrm{RAVE}$.  In the \Rg\ interval 4.5-9.5 kpc, the
$Z_\mathrm{0.0}^\mathrm{RAVE}$ subsample exhibits a gradient of
$\frac{d[Fe/H]}{dR_m}\sim-0.065$~dex, 
which agrees with some of the works cited in Sec.~\ref{intro_grad}.
Moving to the subsamples $Z_\mathrm{0.4}^\mathrm{RAVE}$ and
$Z_\mathrm{0.8}^\mathrm{RAVE}$, the gradients become less negative and then
flatten, as also found in previous works (Cheng et al. \citealp{cheng};
Boeche \citealp{Boeche_PhD}; Pasquali
\& Perinotto \citealp{pasquali}), but with these analyses it is not clear if
the gradient changes smoothly or abruptly.
Thanks to the large number of RAVE stars, we can use a moving box car to
measure the radial gradients along
\Zmax\ (Figure~\ref{Zmax_FeH_grad_bcar}), and show that the gradients (of
four different
elements) are negative and remain approximately constant up to
\Zmax$\sim0.6$~kpc, then they progressively increase  to zero or 
positive values. In the light of the previous discussion of the Besan\c con model,
the reasons for this change may be that i) the kinematically hot component(s) 
have no radial gradients, ii) the fraction of kinematically hot and metal
poor component(s) increase and become significant at \Zmax$>0.6$~kpc,
generating a zero or positive gradient, or iii) both
explanations are valid.\\
The information available up to this point does not allow us to 
decide whether the change of gradient at \Zmax$\sim0.6$~kpc indicates the
transition between one population (thin disc?) to another (thick disc?)
or whether it is a mere artefact. If it were caused by a change from a
population with a gradient (thin disc) to another one without a
gradient (old thin disc or thick disc?) we would also expect a change in
mean abundances at the same \Zmax.
Figure~\ref{Zmax_XH_bcar} shows that the abundances gently
diminish up to \Zmax$\sim1.5$~kpc from the plane. Then, 
the slope becomes steeper until \Zmax$\sim1.8$~kpc, where the abundances become
constant and where one would expect the thick disc to become dominant (but
be aware of the poor statistic between 1.5 and 1.8~kpc). The GCS sample
(Figures~\ref{Zmax_FeH_grad_bcar} and \ref{Zmax_XH_bcar}) exhibits a
similar pattern but at different \Zmax: the rise to a flatter gradient
occurs
between 0.5 and 0.6~kpc and a change of slope in iron abundance at \Zmax$>$
2.0~kpc. 
The change of slope at \Zmax$\sim0.6$~kpc for the gradients and at
\Zmax$\sim1.5$~kpc for the mean abundance is imitated by the abundance 
enhancement [X/Fe] shown in
Figures~\ref{Zmax_XFe_grad_bcar} and \ref{Zmax_XFe_bcar}. 
The stars at \Zmax$<$0.6~kpc and \Zmax$>$1.5~kpc can be identified as thin-
and thick-disc stars, respectively, while the stars in the middle region might be
both the kinematically hot (old) part of the thin disc or a mix of thin- and
thick-disc stars.
The absence of a clear step in abundances at \Zmax$\sim0.6$~kpc
excludes the transition between two populations at that height and supports 
the hypothesis that the change in gradients at \Zmax$\sim0.6$~kpc is caused
by metal-poorer, inner disc (thick disc?) stars, which creates the fictitious flatter
or positive gradient discussed
in Sec.~\ref{sec_mock}. This does not exclude the presence of an old thin
disk population or the absence of a gradient in the thick-disk population
reported by some authors (Allende Prieto et al. \citealp{allende}; Bilir et al.
\citealp{bilir}).\\

\subsection{Observed radial chemical gradients of the thin disc (the
$Z_\mathrm{0.0}^\mathrm{RAVE}$ sample)}
In Sec.~\ref{sec_mock} we discussed that a mix of different stellar 
populations can bias the measured gradients. 
In our data, the stars close to
the Galactic plane are the less affected by this bias. This is confirmed in
Figure~\ref{Zmax_FeH_grad_bcar}, which shows that the gradients remain constant up to
\Zmax$\sim0.6$~kpc because the number of thin-disc stars is larger than any other
stellar populations that might be present in this region. The
$Z_\mathrm{0.0}^\mathrm{RAVE}$ sample is therefore
more suitable for analysing and evaluating the measured chemical gradients
in the framework of the star formation history of the Galactic disc.\\
In the Galactic radius interval 4.5-9.5 kpc, the
$Z_\mathrm{0.0}^\mathrm{RAVE}$ sample
exhibits iron gradients of d[Fe/H]/d$R_g=-0.065$~dex
kpc$^{-1}$, which lies in the range  covered by many observational
studies cited in Sec.~\ref{intro_grad}. The 
alpha elements Mg and Si exhibit similar gradients.
Aluminium has a particularly negative gradient that distinguishes it from 
the other elements, and we briefly discuss it below.\\
Under the assumption of negligible gas infall, the 
abundance [X/H] is a monotonically growing function of the
number of SNe; this means that a negative abundance gradient
implies a higher number of SNe that exploded in the inner disc than
the outer disc in relation to the gas reservoir.
According to this simple remark, chemical evolution models
of the Galaxy with a process of inside-out formation of the disc were
developed
(Chiappini et al. \citealp{chiappini}, \citealp{chiappini2001}; Cescutti el
al. \citealp{cescutti}; Gibson et al. \citealp{gibson}).
Colavitti et al. \cite{colavitti} showed that an 
inside-out formation of the disc is a necessary condition to explain 
the observed metallicity gradient. 
They  predict negative abundance gradients that are slightly flatter than our
values. For instance, Cescutti et al. found 
gradients of --0.052, --0.039 and --0.035~dex kpc$^{-1}$ for Fe, Mg,
and Si, respectively. For comparison, ours are --0.065, --0.073 and --0.064 (see
Table~\ref{tab_XH_grad_Rg}).
Chiappini et al. \cite{chiappini2001} reported a slightly negative gradient in
[$\alpha$/Fe] in the range 4-8~kpc, which turns slighly positive at
8-10~kpc. Gibson et al. \cite{gibson} predicted negative or slightly negative
[O/Fe] gradients at the solar radius. Both works give predicted gradients
close to our findings.\\
The model by Sch\"onrich \& Binney \cite{schoenrich} differs from the
previous ones by construction. This model assumes that the disc forms
simultaneously at all radii (no inside-out formation), with a higher 
infall rate at smaller radii. This causes a higher star formation rate at
small Galactic radii, which generates a negative radial gradient, 
enhanced at late times because the inward
flow of Fe-enriched gas increases the iron abundance (and decreases the ratio
[$\alpha$/Fe]). In addition to these
mechanisms, radial migration constantly acts as flattening agent.
This model predicts an iron gradient of $\simeq$--0.11dex/kpc
and a positive gradient for [$\alpha$/Fe]. This last has the opposite sign
to the sign predicted by the inside-out formation models
and is also opposite to our observational data, which
render a shallow [Mg/Fe] and [Si/Fe] gradient (--0.01 and 0.00,
respectively). This seems to favour the inside-out formation scenario, 
but it does not exclude the flattening action of radial migration 
on the gradients.\\
The comparison with other recent chemo-dynamical models are not conclusive.
The predicted metallicity gradients range from 
--0.02~dex kpc$^{-1}$ (Kobayashi \& Nakasato \citealp{kobayashi}) 
to --0.06~dex kpc$^{-1}$ (Minchev et al. \citealp{minchev}). This last 
gives a Fe gradient that closer agrees with our data, whereas the comparison
with the [Mg/Fe] and [Si/Fe] gradient is not possible.
Indeed, it is not correct to
directly compare the gradients obtained in the present work to the ones in
Figure~5 of Minchev et al. The reason is that in the RAVE analysis our
sample was subdivided in three subsamples as a function of Zmax, hence each
of them composed of stars with a different age mix. As an example, although
in the Figure~5 (bottom) of Minchev et al. the predicted [O/Fe]
gradients are mostly positive, to compute the gradient for a sample composed by
stars with different ages one needs to
take in account the density of stars of different age bins. 
A detailed comparison of the [Mg/Fe] and [Si/Fe] gradients with the 
predictions of the chemodynamical models of
Minchev et al. will be carried out in a forthcoming paper.\\

The aluminium abundance gradient deserves a separate discussion. It is
much more negative ($\sim-0.85$~dex kpc$^{-1}$) than
the other elements. The aluminium abundance suffers from non-LTE effects,
which depend on the stellar parameters and metallicity (Baum\" uller \& Gehren \citealp{baumueller}). 
Although the RAVE chemical abundances are based on LTE spectral analysis,
non-LTE effects are expected to cause mere systematic
errors in our Al abundance estimates based on the Al doublet at
$\sim$8773\AA\
of metal-rich ([m/H]$>-1.0$~dex) dwarf stars. This has no effects on the gradient
determination. Aluminium is also reported to be a useful element to
distinguish different stellar populations, such as disc and halo stars
(Gehren et al. \citealp{gehren2006}). Its peculiar gradient may 
serve as discriminant between different chemical model of the Galaxy.
Unfortunately, these element have not had enough attention in the
cited chemical evolution studies, and a direct comparison with our 
findings is not possible.\\

\section{Conclusions}
We have analysed 19\,962 dwarf stars selected from the RAVE chemical catalogue and
10\,616 dwarf stars selected from the Geneva-Copenhagen survey (GCS) making use of
chemical, kinematic, and Galactic orbit parameters such as chemical abundances,
guiding radii (\Rg), and maximum distance from the Galactic plane
reached by the stars along their orbits (\Zmax).
The same analyses were also applied to a RAVE mock sample created
using the stellar population synthesis code GALAXIA (Sharma et al.
\citealp{sharma}), which is based on the Besan\c con model (Robin et al.
\citealp{robin}). 
Using the RAVE and GCS sample, we found that the Galactic disc has a
negative iron abundance gradient of $-0.065$~dex~kpc$^{-1}$ and
$-0.043$~dex~kpc$^{-1}$ for the RAVE and GCS samples, respectively. 
We found that for the GCS sample the gradient is more negative
($-0.054$~dex~kpc$^{-1}$) when
stars at small \Rg\ ($<5.5$~kpc) are excluded from the measurement.
This is due to an observational bias that affects volume-limited 
samples such as the GCS and RAVE:  stars from the inner disc can reach the
Sun's neighbourhood only if they move on eccentric orbits.  Such
kinematically hot stars are on average more metal-poor than stars with the same
\Rg, but are on more circular orbits, which will be missing from our sample.  This
creates a bias against high-metallicity and low-\Rg\ stars, which contributes to
make the gradient flatter than it really is. This effect is evident in the 
GCS sample and barely visible in the RAVE one. To assess the impact
of this bias, we measured the gradients with different cuts at low and high
\Rg, with differences in gradients no larger than 0.01~dex~kpc$^{-1}$.
This uncertainty in the chemical gradients is small enough not to affect 
our data interpretation. The presence of inhomogeneities in the 
volumes probed by the RAVE and GCS samples (such as disrupted open clusters
and moving groups) cannot be excluded to
explain the different obtained gradients. Despite the differences just
highlighted, for both RAVE and GCS samples the gradient shows a common
trend: it becomes flatter at
increasing \Zmax\ (as found by previous studies, e.g.  Cheng et al. 
\citealp{cheng}; Boeche \citealp{Boeche_PhD}; Pasquali \& Perinotto
\citealp{pasquali}).  The RAVE sample also allows gradient measurements
for the elements Mg, Al, and Si, for which we found the same trend.

The GALAXIA/Beson\c con mock sample shows striking differences compared
with the RAVE and GCS samples, because the iron gradients are positive at
any \Zmax. This surprising and unrealistic result is caused by the lack of
correlation between metallicity and tangential velocity (or angular 
momentum) in the Besan\c con model. In fact, although the Besan\c con model
assumes a radial gradient of --0.07~dex~kpc$^{-1}$ for the thin disc (and
zero for the other populations), the metallicities are assigned regardless
of the kinematics. This means that stars with small \Rg\ have the same
probability to be metal-rich as stars with large \Rg, while in the
real Galaxy stars coming from the inner disc are more likely to be 
metal-rich than stars coming from the outer disc. The absence of a correlation
between chemistry and kinematics in the Besan\c con model removes the
metallicity gradient in the mock sample when the kinematically dependent
\Rg\ quantity is used. Moreover, the thick-disc stars, with a mean metallicity of
$-0.78$~dex and stronger asymmetric drift than the thin-disc stars, shift to
lower \Rg\ and [Fe/H], creating a positive metallicity gradient 
in the full sample even when the
individual components have a negative or no radial gradient. 
The mock sample also shows an excess of
thick disc stars compared with the RAVE (and GCS) sample, with a mean
metallicity peak at $-0.78$~dex. This appears too low compared with
the RAVE and GCS samples, where a value of $-0.5$~dex seems more realistic
(also suggested by \citealp[Bilir et al.][]{bilir};
\citealp[Kordopatis et al.][]{kordopatis_b}; \citealp[Schlesinger et
al.][]{schlesinger}; \citealp[Soubiran et al.][]{soubiran03}). The
discrepancies between the observational samples and the model can be reduced
by i) decreasing the local density, ii) decreasing the vertical
velocity dispersion, and iii) raising the mean metallicity of the thick disc.\\
Our analysis of the gradients as function of \Zmax\ with a moving box car
revealed that the abundance gradients remain approximately constant up to $\sim$0.6~kpc, and
then increase to zero. The change in gradients at $\sim$0.6~kpc does not
correspond to a change in average abundance, suggesting that the gradients
become flatter because the fraction of kinematically hot, metal-poor 
(thick-disc) stars become significant at that height and leads to a
flatter gradient, as discussed before. Another explanation is that this
population of stars has no radial gradient. Both can be true, and our 
analysis is not able to distinguish between these possibilities.
The constancy of the radial gradients up to \Zmax$\sim$0.6~kpc reveals that
a stellar population with gradient zero (probably the thick
disc) does not affect the thin disc properties
at these heights and the values obtained here
are reliable. For a sample of RAVE stars at \Zmax$<$0.4~kpc we measured
gradients that are steeper (d[Fe/H]/d\Rg=--0.065 dex kpc$^{-1}$)
than predicted by some chemical models (for instance --0.040 dex kpc$^{-1}$
by Chiappini et al. \citealp{chiappini2001}, --0.052 dex kpc$^{-1}$ by
Cescutti et al. \citealp{cescutti}) and flatter than others 
(--0.11 dex kpc$^{-1}$ by Sch\"onrich \& Binney \citealp{schoenrich}).
We also found slightly negative (or zero) gradients for abundances relative to iron
(i.e. d[Mg/Fe]/d\Rg=--0.009 dex kpc$^{-1}$), which supports
the models of Chiappini et al. \cite{chiappini2001} and Gibson et al.
\cite{gibson}, and it is in contrast with 
the model by Sch\"onrich \& Binney \cite{schoenrich} which predict positive
gradients.\\
The last model considers radial migration 
to be an important flattening agent for the radial
gradients. Although it is common sense to expect that radial mixing is
acting in our Galaxy, it is difficult to estimate the impact of its action
on the observed chemical gradients, because migrated stars changed their
original \Rg\ (Ro{\v s}kar et al. \citealp{roskar08};
S{\'a}nchez-Bl{\'a}zquez et al. \citealp{sanchez}) and have become kinematically 
indistinguishable from the local sample. Detailed chemical analyses are
therefore necessary to distinguish migrated from the locally
born stars. High-resolution and high signal-to-noise spectroscopic surveys
such as the Gaia-ESO survey \cite[Gilmore et al.][]{gilmore}, 
the GALactic Archaeology with HERMES (GALAH) survey \cite[Zucker et
al.][]{zucker}, 
and the Apache Point Observatory Galactic Evolution Experiment (APOGEE)
survey \cite[Majewski et al.][]{majewski} may be able to identify 
and quantify migrated stars.

This paper will be followed by a second one in which we measure the radial
chemical gradients by using RAVE giant stars. Because RAVE is a magnitude-limited survey,
and because giants stars are luminous objects, we are able to cover a larger
volume, use the actual Galactocentric distance $R$ in addition to the guiding
radius \Rg, and probe the Galactic disc at larger heights.
This analysis will provide additional constraints on the chemical and kinematical
characteristics of the Milky Way.

\begin{acknowledgements}

We acknowledge funding from Sonderforschungsbereich SFB 881 ``The Milky Way
System" (subproject A5) of the German Research Foundation (DFG).  Funding
for RAVE has been provided by the Australian Astronomical Observatory; the
Leibniz-Institut fuer Astrophysik Potsdam (AIP); the Australian National
University; the Australian Research Council; the French National Research
Agency; the German Research Foundation (SPP 1177 and SFB 881); the European
Research Council (ERC-StG 240271 Galactica); the Istituto Nazionale di
Astrofisica at Padova; The Johns Hopkins University; the National Science
Foundation of the USA (AST-0908326); the W.  M.  Keck foundation; the
Macquarie University; the Netherlands Research School for Astronomy; the
Natural Sciences and Engineering Research Council of Canada; the Slovenian
Research Agency; the Swiss National Science Foundation; the Science \&
Technology Facilities Council of the UK; Opticon; Strasbourg Observatory;
and the Universities of Groningen, Heidelberg and Sydney.  The RAVE web site
is at http://www.rave-survey.org.

\end{acknowledgements}


\begin{thebibliography}{}
%
\bibitem[2006]{allende} Allende Prieto, 
C., Beers, T.~C., Wilhelm, R., et al.\ 2006, \apj, 636, 804 
%
\bibitem[2002]{andrievsky} Andrievsky, S.~M., Kovtyukh, V.~V., Luck,
R.~E., et al.\ 2002, \aap, 381, 32 
%
\bibitem[2012]{antoja} Antoja, T., Helmi, A., 
Bienayme, O., et al.\ 2012, \mnras, 426, L1  
%
\bibitem[1997]{baumueller} Baum{\"u}ller, D., \& Gehren, T.,\ 1997, \aap,
325, 1088
%
\bibitem[2012]{bilir} Bilir, S., Karaali, S., 
Ak, S., et al.\ 2012, \mnras, 421, 3362 
%
\bibitem[2010]{binney2010} Binney, J.\ 2010, \mnras, 401, 
2318 
%
\bibitem[2013]{binney2013} Binney, J., Burnett, B., Kordopatis, G., et al.,
2013, submitted
%
\bibitem[2013]{boeche2013} Boeche, C., Chiappini, C., Minchev, I., et
al.\ 2013, \aap, 553, A19
%
\bibitem[2011a]{Boeche_PhD} Boeche, C., PhD thesis, 2011a,
	urn:nbn:de:kobv:517-opus-52478, Potsdam
	Universit\"at, url {http://opus.kobv.de/ubp/volltexte/2011/5247/}
%
\bibitem[2011b]{boeche} Boeche, C., Siebert, A., Williams, M., et al., 2011b,
\aj, 142, 193
%
\bibitem[2010]{breddels} Breddels, M.~A., Smith, M.~C., Helmi, A., et
al.\ 2010, \aap, 511, A90 
%
%
\bibitem[2011]{burnett} Burnett, B., Binney, J., Sharma, S., et al.\
2011, \aap, 532, A113 
%
\bibitem[2007]{carraro} Carraro, G., Geisler, D., Villanova, S.,
Frinchaboy, P.~M., \& Majewski, S.~R.\ 2007, \aap, 476, 217 
%
\bibitem[2012]{carrell} Carrell, K., Chen, Y., 
\& Zhao, G.\ 2012, \aj, 144, 185 
%
\bibitem[2010]{casagrande10} Casagrande, L., Ram{\'{\i}}rez, I.,
Mel{\'e}ndez, J., Bessell, M., \& Asplund, M.\ 2010, \aap, 512, A54 
%
\bibitem[2011]{casagrande} Casagrande, L., Sch\"onrich, R., Asplund, M.,
Cassisi, S., Ramirez, I., Mel\'endez, J., Bensby, T., Feltzing, S., 2011,
A\&A 530, 138
%
\bibitem[2007]{cescutti} Cescutti, G., Matteucci, F., Fran{\c c}ois, P., \&
Chiappini, C.\ 2007, \aap, 462, 943
%
\bibitem[2012]{cheng} Cheng, J.~Y., Rockosi, 
C.~M., Morrison, H.~L., et al.\ 2012, \apj, 746, 149 
%
\bibitem[1997]{chiappini} Chiappini, C., 
Matteucci, F., \& Gratton, R.\ 1997, \apj, 477, 765 
%
\bibitem[2001]{chiappini2001} Chiappini, C., 
Matteucci, F., \& Romano, D.\ 2001, \apj, 554, 1044 
%
\bibitem[2009]{colavitti} Colavitti, E., Cescutti, G., Matteucci, F.,
\& Murante, G.\ 2009, \aap, 496, 429 
%
\bibitem[2012]{coskunoglu} Co{\c 
s}kuno{\v g}lu, B., Ak, S., Bilir, S., et al.\ 2012, \mnras, 419, 2844 
%
\bibitem[1998]{dehnen} Dehnen, W., Binney, J.\ 1998, \mnras, 294, 429
%
\bibitem[2005]{famaey} Famaey, B., Jorissen, A., Luri, X., et al.\
2005, \aap, 430, 165 
%
\bibitem[2002]{friel} Friel, E.~D., Janes, 
K.~A., Tavarez, M., et al.\ 2002, \aj, 124, 2693
%
%
\bibitem[2006]{gehren2006} Gehren, T., Shi, J.~R., Zhang, H.~W., Zhao,
G., \& Korn, A.~J.\ 2006, \aap, 451, 1065 
%
\bibitem[2013]{gibson} Gibson, B.~K., Pilkington, K., Brook, C.~B.,
Stinson, G.~S., \& Bailin, J.\ 2013, \aap, 554, A47 
%
\bibitem[2012]{gilmore} Gilmore, G., Randich, 
S., Asplund, M., et al.\ 2012, The Messenger, 147, 25 
%
\bibitem[2005]{gorski} G{\'o}rski, K.~M., 
Hivon, E., Banday, A.~J., et al.\ 2005, \apj, 622, 759 
\bibitem[2000]{hog} H{\o}g, E., Fabricius, C., Makarov, V.~V., et al.\ 2000,
 \aap, 355, L27 
%
%
\bibitem[2008]{ivezic} Ivezi{\'c}, {\v Z}., 
Sesar, B., Juri{\'c}, M., et al.\ 2008, \apj, 684, 287
%
\bibitem[2008]{juric} Juri{\'c}, M., 
Ivezi{\'c}, {\v Z}., Brooks, A., et al.\ 2008, \apj, 673, 864 
%
\bibitem[2012]{karatas} Karata{\c s}, Y., \& Klement, R.~J.\
2012, \na, 17, 22 \
%
\bibitem[2011]{katz} Katz, D., Soubiran, C., Cayrel, R., et al.\
2011, \aap, 525, A90 
%
\bibitem[2011]{kobayashi} Kobayashi, C., \& Nakasato, N.\
2011, \apj, 729, 16 
%
\bibitem[2011a]{kordopatis} Kordopatis, G., Recio-Blanco, A., de
Laverny, P., et al.\ 2011, \aap, 535, A106 
%
\bibitem[2011b]{kordopatis_b} Kordopatis, G., Recio-Blanco, A., de
Laverny, P., et al.\ 2011, \aap, 535, A107 
%
\bibitem[2013]{kordopatis13} Kordopatis, G., Gilmore, G., Steinmetz, M., et
al.\ 2013, \aj accepted
%
\bibitem[2008]{lemasle} Lemasle, B., Fran{\c c}ois, P., Piersimoni,
A., et al.\ 2008, \aap, 490, 613
%
%
\bibitem[2006]{luck} Luck, R.~E., Kovtyukh, 
V.~V., \& Andrievsky, S.~M.\ 2006, \aj, 132, 902 
%
\bibitem[2011]{luck2011} Luck, R.~E., \& Lambert, D.~L.\ 2011,
\aj, 142, 136 
%
\bibitem[2010]{maciel2010} Maciel, W.~J., \& Costa, R.~D.~D.\
2010, IAU Symposium, 265, 317 
%
\bibitem[1999]{maciel} Maciel, W.~J., \& Quireza, C.\ 1999,
\aap, 345, 629 
%
\bibitem[2010]{majewski} Majewski, S.~R., 
Wilson, J.~C., Hearty, F., Schiavon, R.~R., 
\& Skrutskie, M.~F.\ 2010, IAU Symposium, 265, 480 
%
\bibitem[2012]{matijevic} Matijevi{\v c}, 
G., Zwitter, T., Bienaym{\'e}, O., et al.\ 2012, \apjs, 200, 14 
%
\bibitem[1989]{matteucci} Matteucci, F., \& Francois, P.\ 1989, \mnras, 239,
885 
%
\bibitem[2011a]{minchev11a} Minchev, I., Famaey, B., Combes, F., et al.\ 2011, \aap, 527, A147
%
\bibitem[2012]{minchev} Minchev, I., Chiappini, 
C., \& Martig, M.\ 2012, arXiv:1208.1506, in press 
%
\bibitem[2004]{nordstrom} Nordstr\"om, B., Mayor, M., Andersen, J., et al.\ 2004, A\&A 418, 989
%
\bibitem[2010]{pancino} Pancino, E., Carrera, R., Rossetti, E., \&
Gallart, C.\ 2010, \aap, 511, A56 
%
\bibitem[2012a]{pasetto_a} Pasetto, S., Grebel, E.~K., Zwitter, T., et
al.\ 2012a, \aap, 547, A70 
%
\bibitem[2012b]{pasetto_b} Pasetto, S., Grebel, E.~K., Zwitter, T., et
al.\ 2012b, \aap, 547, A71
%
\bibitem[1993]{pasquali} Pasquali, A., Perinotto, M.\ 1993, A\&A 280, 581
%
\bibitem[2012]{pilkington} Pilkington, K., Few, C.~G., Gibson, B.~K.,
et al.\ 2012, \aap, 540, A56 
%
%
\bibitem[2003]{robin} Robin, A.~C., Reyl{\'e}, C., Derri{\`e}re,
S., \& Picaud, S.\ 2003, \aap, 409, 523 
%
\bibitem[2008]{roeser2008} R{\"o}ser, S., Schilbach, E., Schwan, H., et
al.\ 2008, \aap, 488, 401 
%
\bibitem[2010]{roeser2010} Roeser, S., Demleitner, 
M., \& Schilbach, E.\ 2010, \aj, 139, 2440 
%
\bibitem[2008]{roskar08} Ro{\v s}kar, R., 
Debattista, V.~P., Quinn, T.~R., Stinson, G.~S., 
\& Wadsley, J.\ 2008, \apjl, 684, L79
%
\bibitem[2011]{ruchti} Ruchti, G.~R., 
Fulbright, J.~P., Wyse, R.~F.~G., et al.\ 2011, \apj, 737, 9 
%
\bibitem[2009]{sanchez} 
S{\'a}nchez-Bl{\'a}zquez, P., Courty, S., Gibson, B.~K., 
\& Brook, C.~B.\ 2009, \mnras, 398, 591 
%
\bibitem[2012]{schlesinger} Schlesinger, K.~J., 
Johnson, J.~A., Rockosi, C.~M., et al.\ 2012, \apj, 761, 160 
%
\bibitem[2009a]{schoenrich} Sch{\"o}nrich, R., \& Binney, J.\
2009, \mnras, 396, 203 
%
\bibitem[2009b]{schoenrich_b} Sch{\"o}nrich, R., \& Binney, J.\
2009, \mnras, 399, 1145 
%
%
\bibitem[2002]{sellwood02} Sellwood, J.~A., \& Binney, J.~J.\
2002, \mnras, 336, 785 
%
\bibitem[2008]{sestito} Sestito, P., Bragaglia, A., Randich, S., et
al.\ 2008, \aap, 488, 943 
%
\bibitem[2011]{sharma} Sharma, S., 
Bland-Hawthorn, J., Johnston, K.~V., \& Binney, J.\ 2011, \apj, 730, 3 
%
\bibitem[2011]{siebert} Siebert, A., Williams, M.~E.~K., Siviero, A., et
al.\ 2011, \aj, 141, 187
%
\bibitem[2011]{siebert2011} Siebert, A., Famaey, 
B., Minchev, I., et al.\ 2011, \mnras, 412, 2026 
%
\bibitem[2003]{soubiran03} Soubiran, C., Bienaym\'e, O., Siebert, A.\ 2003,
A\&A, 398, 141
%
\bibitem[1995]{teuben} Teuben, P.J., The Stellar Dynamics Toolbox NEMO, in: 
Astronomical Data Analysis Software and Systems IV, 
ed. R. Shaw, H.E. Payne and J.J.E. Hayes. (1995),
PASP Conf Series 77, 398
%
\bibitem[2006]{rave} Steinmetz, M., Zwitter, T., Siebert, A., et al.\ 2006,
\aj, 132, 1645
%
%
\bibitem[2008]{veltz} Veltz, L., Bienaym{\'e}, O., Freeman, K.~C.,
et al.\ 2008, \aap, 480, 753
%
\bibitem[2013]{williams} Williams, M.~E.~K., 
Steinmetz, M., Binney, J., et al.\ 2013, arXiv:1302.2468 
%
\bibitem[2011]{wilson} Wilson, M.~L., Helmi, 
A., Morrison, H.~L., et al.\ 2011, \mnras, 413, 2235
%
\bibitem[2006]{yong} Yong, D., Carney, B.~W., 
Teixera de Almeida, M.~L., \& Pohl, B.~L.\ 2006, \aj, 131, 2256 
%
\bibitem[2012]{yong2012} Yong, D., Carney, B.~W., 
\& Friel, E.~D.\ 2012, \aj, 144, 95 
%
\bibitem[2004]{zacharias} Zacharias, N., Urban, S.~E., Zacharias, M.~I., 
et al.\ 2004, \aj, 127, 3043
%
\bibitem[2012]{zucker} Zucker, D.~B., de Silva, 
G., Freeman, K., Bland-Hawthorn, J., 
\& Hermes Team 2012, Galactic Archaeology: Near-Field Cosmology and the
Formation of the Milky Way, 458, 421 
%
\bibitem[2008]{zwitter} Zwitter, T., Siebert, A., Munari, U., et al.\ 2008,
\aj, 136, 421
%
\bibitem[2010]{zwitter2010} Zwitter, T., Matijevi\v{c}, G., Breddels, M. A.,
et al.\ 2010, A\&A, 522, 54
\end{thebibliography}
\end{document}